\documentclass[%
reprint,
pre,
superscriptaddress,
onecolumn,
nofootinbib,
 amsmath,amssymb,
 aps,
]{revtex4-2}

\usepackage{graphicx}
\usepackage{dcolumn}
\usepackage{bm}
\usepackage{xcolor}
\usepackage{hyperref}
\hypersetup{
    pdfstartview={FitH},    
    colorlinks=true,       
    linkcolor=blue,          
    citecolor=blue,        
    filecolor=magenta,      
    urlcolor=blue           
}

\begin{document}
\linespread{1.05}


\title{Nanoscale sensing of spatial correlations in nonequilibrium current noise}

\author{Yifan F. Zhang}
\email{yz4281@princeton.edu}
\affiliation{%
 Department of Electrical and Computer Engineering, Princeton University, Princeton, NJ 08544, USA
}%
\author{Rhine Samajdar}
\email{rhine\_samajdar@princeton.edu}
\affiliation{Department of Physics, Princeton University, Princeton, NJ 08544, USA}
\affiliation{Princeton Center for Theoretical Science, Princeton University, Princeton, NJ 08544}
 \author{Sarang Gopalakrishnan}
\affiliation{%
 Department of Electrical and Computer Engineering, Princeton University, Princeton, NJ 08544, USA
}%

\begin{abstract}
Nitrogen-vacancy centers are spatially resolved probes of current noise. 
So far, current noise sensing with NV centers has primarily been used as a way to probe equilibrium transport coefficients. We develop a framework for computing the spatiotemporal correlations of nonequilibrium current noise in the Boltzmann regime, and apply it to two-dimensional metals in current-biased steady states. 
We argue that the spatial structure of the noise reveals the nonequilibrium nature of the electron distribution function, and more generally reveals the nature and lifetimes of the excitations responsible for transport. 
We estimate the visibility of these signatures in near-term experiments.
\end{abstract}

\maketitle

\section{Introduction}

Noise in the current (or other local observables) in a material carries a wealth of information about material properties. Equilibrium current noise is related to the conductivity by the fluctuation-dissipation theorem, and is a powerful noninvasive probe of properties like the conductivity or the temperature~\cite{qu2019johnson, Kolkowitz2015a, PhysRevB.95.155107}. Away from equilibrium, response functions and noise carry distinct information; for instance, noise probes 
can reveal information beyond linear response, like
the charge of individual quasiparticles~\cite{de1998direct, reznikov1999observation, PhysRevLett.72.724, PhysRevB.51.2363, PhysRevLett.75.2196, PhysRevB.95.115308}.  
It was recently realized that the nature of nonequilibrium ``shot'' noise in strange metals without well-defined quasiparticles~\cite{wang2022shot, chen2023shot, PhysRevResearch.5.043143, wu2023suppression} (and near quantum critical points~\cite{PhysRevLett.97.227003}) can strongly constrain the theoretical explanations of their transport. Moreover, even far from equilibrium, nonequilibrium noise can clearly distinguish between qualitatively different mechanisms for diffusion~\cite{wei2022quantum, PhysRevLett.128.090604, PhysRevLett.128.160601, PhysRevB.109.024417, PhysRevLett.131.210402, rosenberg2023dynamics, wienand2023emergence, samajdar2023quantum, gopalakrishnan2024non}. 
Experimentally, nonequilibrium shot noise has also been widely studied in metal nanostructures \cite{blanter2000shot,tikhonov2020spatial}, cold atom gases \cite{wienand2023emergence}, and superconducting devices \cite{rosenberg2023dynamics}.
Most of these works have focused on current fluctuations across a single point in a sample, typically a constriction if the sample is not one-dimensional. Nonequilibrium noise in the ``bulk'' of a two-dimensional sample, and the spatial correlations of noise, have been much less explored, as these quantities are not accessible to most standard ways of probing noise. 

Over the past decade, nitrogen-vacancy (NV) centers in diamond have emerged as sensitive local probes of current noise~\cite{Kolkowitz2015a, Pelliccione2016a, Ariyaratne2018a, PhysRevB.98.195433, Andersen2019a, Lee-Wong2020, 10.1038/s41586-020-2507-2, Zhou2021a,10.1038/s41567-021-01341-w, 10.1103/physrevlett.129.087701, Rovny2022a, rovny2024new}. An NV center is effectively a qubit precessing in a local field generated by currents in the underlying sample; thus, the decoherence of the NV center can be used to diagnose the noise power spectrum. 
Being solid-state defects, NV centers can be brought as close to a sample as $10$ nm, providing unprecedented spatial resolution.
%
In the frequency domain, depending on the sensing protocol, they can readily sense fluctuations in a range from $\alt 1$ MHz to $\sim 1$ GHz. 
A recent advance in NV sensing was the development of a covariance magnetometry protocol, which enables sensing two-point correlations of noise across $\sim $ 1 $\mu $m length scales by measuring two NV centers simultaneously~\cite{Rovny2022a}, opening up new possibilities in sensing spatially correlated nonequilibrium fluctuations. Recent advances in scalable parallel readout of large NV arrays~\cite{cheng2025massively,cambria2025scalable} further extend these capabilities and motivate the present theoretical study.

In this work, we explore nonequilibrium noise in current-driven metals using covariance magnetometry. While previous experimental and theoretical applications of NV covariance magnetometry have focused predominantly on equilibrium phenomena (e.g., Johnson noise, magnetic textures, and thermal transport coefficients), here we show that the same observable, in a current-biased steady state, contains qualitatively different and richer information that is invisible to single-NV magnetometry and to standard global transport measurements. Specifically, we compute spatiotemporal noise correlations in the Boltzmann regime and predict strongly spatially anisotropic current noise in the nonequilibrium steady state. The spatial anisotropy is a hallmark of the nonequilibrium nature of the steady state and is detectable with covariance magnetometry. The direction of enhancement reveals the underlying charge carriers, e.g., whether they are electrons or magnetic vortices, while the extent of this enhancement reveals the relative rates of energy and momentum relaxation. We estimate the visibility of these noise signatures in near-term covariance magnetometry experiments.

\section{Experimental setup}

\begin{figure}[h]
\center
\includegraphics[width=0.5\linewidth]{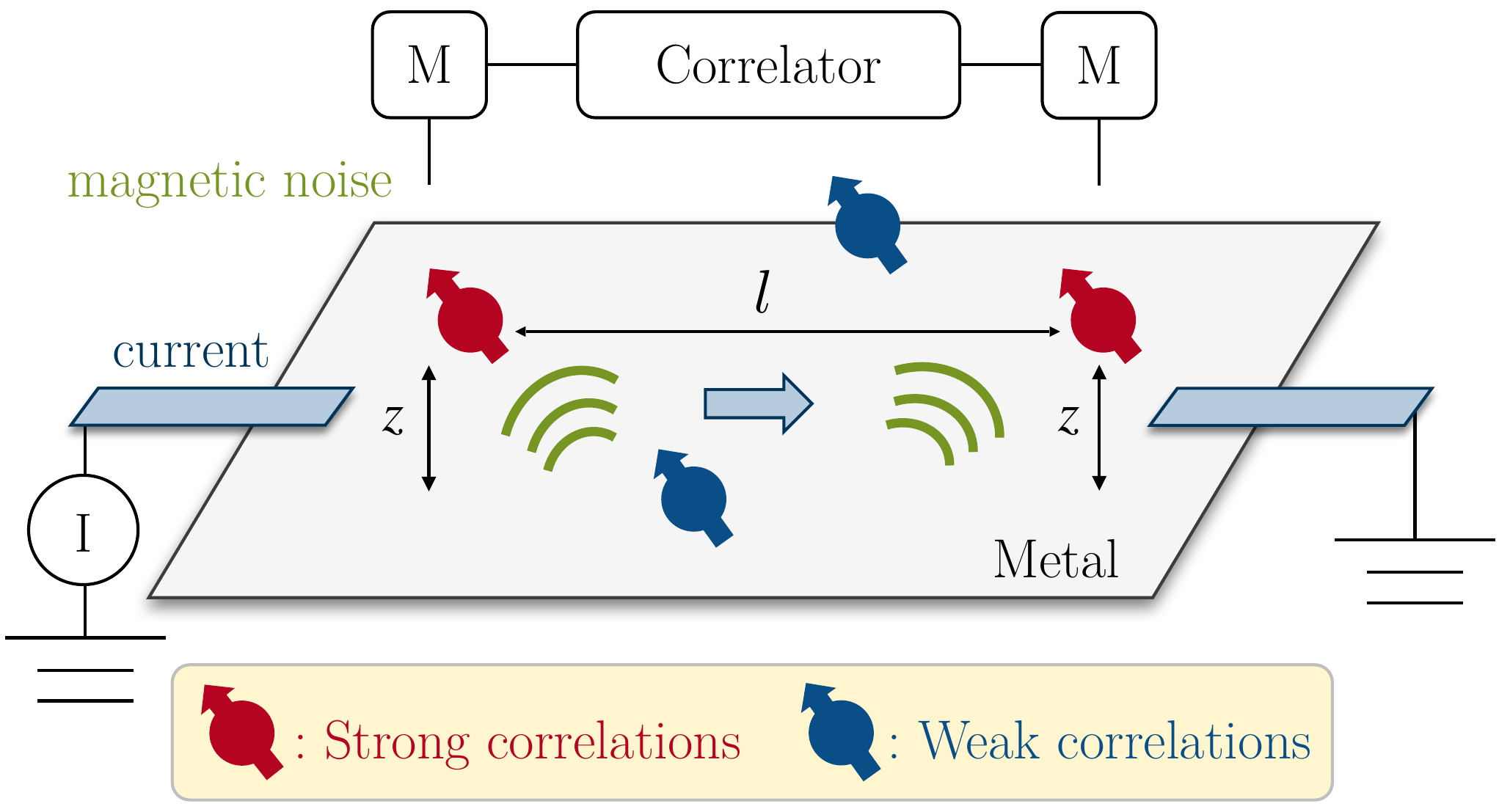}
\caption{\label{fig:setup}Setup for detecting the spatial correlations of current noise using NV centers separated by a distance $l$ and situated at a height $z$ above a two-dimensional sample, in the presence of an external current. In the Boltzmann regime analyzed below, noise along the current flow is stronger than noise perpendicular to the flow.}
\end{figure}

We first outline the experimental setup, shown in Fig.~\ref{fig:setup}. A set of individually addressable NV centers is placed near the surface of the metal at depth $z$. The depth can vary between NV centers, but—as we show later—the covariance signal is insensitive to this variation, so these fluctuations can be ignored. A current is driven through the metal to generate nonequilibrium noise. We then perform covariance magnetometry between pairs of NV centers aligned either parallel or perpendicular to the current direction. In practice, all NV centers can be measured simultaneously, and the pairwise correlations can be computed in post-processing~\cite{cheng2025massively,cambria2025scalable}.

We show that a clear signature of nonequilibrium noise is its \emph{anisotropy}: NV pairs aligned parallel to the current direction experience a stronger correlated noise signal than NV pairs aligned perpendicular to it. This contrast does not arise from simple heating, which increases noise in an isotropic manner.\\

\section{Methods}

\subsection{Boltzmann formalism}

We will focus on transport due to quasiparticles in a two-dimensional metal. 
We begin with the semiclassical Boltzmann equation for the electron occupation function:
\begin{align}\label{boltzmann}
    \partial_t f(\vec{r},\vec{k},t)&=-(\vec{v}_{\vec k} \cdot \nabla_r + \vec{F}_{\mathrm{ext}} \cdot \nabla_k) f - I_{\mathrm{col}}[f],
\end{align}
where $\vec{v}_{\vec k}$ is the momentum-dependent group velocity, $\vec{F}_{\mathrm{ext}}$ is the external field, and $I_{\mathrm{col}}[f]$ is the collision functional,
\begin{equation}
\begin{split}
      I_{\mathrm{col}}[f] = \sum_{\Delta \vec{k}}[&W_{\Delta \vec{k}}f(\vec{r},\vec{k},t)(1-f(\vec{r},\vec{k}+\Delta\vec{k},t))
    -W_{-\Delta \vec{k}}f(\vec{r},\vec{k}+\Delta\vec{k},t)(1-f(\vec{r},\vec{k},t))]
    \end{split}
\end{equation}
with $W_{\Delta \vec{k}}$ denoting the scattering rate from the mode $\vec{k}$ to $\vec{k} + \Delta \vec k$. 
We will work in a semiclassical, high-conductance regime, featuring a separation of scales between the Fermi momentum $k_F$ and the typical scales on which the distribution varies, which we take to be comparable to or larger than the mean free path. 
All our observables will be coarse-grained over a length scale $\gg k_F^{-1}$, so that mesoscopic coherence effects~\cite{PhysRevLett.123.110601, PhysRevX.13.011045} are washed out. 

We take the occupation function at each $(\vec{r}, \vec{k}, t)$ to be a random variable $\hat f (\vec{r}, \vec{k}, t)$. The mean of this random variable is $\langle \hat f \rangle = f_{\mathrm{ss}}$, the steady-state distribution function, which annihilates Eq.~\eqref{boltzmann}.
{Assuming that the equal-time fluctuations in $\hat f$ are short-range correlated, the steady state is fully specified by $f_{\mathrm{ss}}$}. Under this standard assumption~\cite{kogan1969theory, PhysRev.187.267, PhysRevB.46.1889,gantsevich1979theory}, the variance of $\hat f$ is given by the equal-time correlator $f_2(\vec{r},\vec{k},t; \vec{r}', \vec{k}', t) = \delta(\vec{r} - \vec{r}') \delta(\vec{k} - \vec{k}') \varphi(\vec{r}, \vec{k})$, which is assumed to be uncorrelated on the scales we are considering. Finally, since each $\vec{k}$-mode is either occupied or empty, $\varphi$ follows a Bernoulli distribution, so 
\begin{equation}\label{equaltime}
f_2(\vec{r},\vec{k},t;\vec{r'},\vec{k'},t)\!=\!\delta(\vec{r}-\vec{r'})\delta(\vec{k}-\vec{k'})f_{\mathrm{ss}}(\vec{r},\vec{k})(1-f_{\mathrm{ss}}(\vec{r},\vec{k})).
\end{equation}
An important observation~\cite{PhysRev.187.267,kogan1969theory,gantsevich1979theory} is that the unequal-time correlation function $f_2(\vec{r},\vec{k},t+\tau;\vec{r'},\vec{k'},t)$ evolves precisely according to the deterministic Boltzmann equation~\eqref{boltzmann}, where the time derivative is taken with respect to $\tau$~\footnote{This follows, e.g., from the equation relating the Keldysh and retarded Green's functions, together with the observation that the distribution function is stationary.}. The intuition is that an initial density fluctuation is advected and spread out by precisely the same processes as those captured by Eq.~\eqref{boltzmann}. Thus, to understand the evolution of unequal-time correlations, one simply has to solve Eq.~\eqref{boltzmann} with the initial condition being uncorrelated Bernoulli noise~\eqref{equaltime}. 

\subsection{Relaxation-time approximation}

As a first application of these ideas, we take the simplest collision functional, namely that given by the relaxation-time approximation (RTA)~\cite{pitaevskii2012physical}:
$I_{\mathrm{col}}[f] \approx (f-f_{\mathrm{th}})/\tau$, where $f_{\mathrm{th}}$ is the thermal distribution and $\tau$ is the scattering time\footnote{See Appendix~\ref{sec:app} for additional discussions beyond RTA.}. In this approximation, any scattering event totally randomizes $\vec{k}$ and decorrelates the current. Hence, we can treat $f_2$ as remaining diagonal at all times and evolving according to the collisionless Boltzmann equation augmented with a decay term: 
\begin{equation} \label{nonlocal_scattering_f2}
    \partial_{t} f_2(\vec{r},\vec{k},t) =
-(\vec{v}_{\vec k} \cdot \nabla_r + \vec{F}_{\mathrm{ext}} \cdot \nabla_k) f_2 - \frac{ f_2}{\tau}
\end{equation}
This approximation neglects the sum rule for $f_2$ stemming from the conservation law. These hydrodynamic correlations would be crucial if we were probing the density noise, but are unimportant for current-current correlations except potentially at the longest length scales (due to hydrodynamic long-time tails~\cite{forster2018hydrodynamic}).

\section{NV magnetometry}

In the Boltzmann formalism, the instantaneous current due to mode $\vec{k}$ is $q\,\vec{v}_{\vec k} \hat f(\vec{k})$. Thus we can express the current correlation function in terms of $f_2$ as:
\begin{equation}\label{jjcorr}
\begin{split}
    \langle \vec{J}(\vec{r},t+\Delta t)\vec{J}(\vec{r}',t)\rangle -\langle \vec{J}(\vec{r},t+\Delta t) \rangle \langle \vec{J}(\vec{r}',t) \rangle=
    \int d\vec{k} d\vec{k}' q^2 \vec{v}_{\vec k} \vec{v}_{\vec k'} f_2(\vec{r},\vec{k},t+\Delta t;\vec{r'},\vec{k'},t),
    \end{split}
\end{equation}
where $q$ represents the electron charge.
The magnetic field at the location of the NV center is related to the current~\cite{rovny2024new} by the Biot-Savart law
\begin{equation} \label{biot_savart}
    \vec{B}(\vec{\rho},t)=\frac{\mu_0}{4 \pi}\int d \vec{r}\, \frac{\vec{J}(\vec{r},t) \times (\vec{\rho}-\vec{r})}{|\vec{\rho}-\vec{r}|^3},
\end{equation}
where {$\mu_0$ is the vacuum permeability,} and $\vec{\rho}$ is a three-dimensional vector with the out-of-plane component being the NV depth $z$ (contrary to $\vec{r}$, which does not have an out-of-plane component). In the case of magnetometry with a single NV, we evaluate the (connected) magnetic fluctuations at the same location $\langle \vec{B}(\vec{\rho},t+\Delta t)\vec{B}(\vec{\rho},t)\rangle$; for covariance magnetometry,  we evaluate the (connected) magnetic fluctuations at the two NV centers $\langle \vec{B}(\vec{\rho}^{}_{\textsc{nv}1},t+\Delta t)\vec{B}(\vec{\rho}^{}_{\textsc{nv}2},t)\rangle$.

Finally, the physical {observable} is the phase noise: $\langle \phi^2 \rangle$ with single-NV magnetometry and $\langle \phi^{}_{\textsc{nv}1} \phi^{}_{\textsc{nv}2} \rangle$ in the case of covariance magnetometry. The phase is connected to the spin dephasing over time under a magnetic field,
\begin{equation}\label{nvnoise}
    \phi(T) = \gamma \int_{0}^{T} dt \; \hat{n} \cdot \vec{B}(\vec{\rho},t);
\end{equation}
$\gamma = 2\pi \times 28$~GHz/T is the gyromagnetic ratio, $\hat{n}$ is a unit vector pointing along the NV orientation, and the phase accumulation can be used for sensing only on timescales $T \leq T_2$, where $T_2$ is the intrinsic dephasing time of the NV center.
The experimental protocol for correlated sensing~\cite{Rovny2022a} in effect extracts the correlation function $\langle \exp[i (\phi_1(T) - \phi_2(T))]\rangle$, where $\phi_i(T)$ is the phase picked up by the $i$th NV. 
For Gaussian noise, one can rewrite this observable as $\exp(-\frac{1}{2} \langle(\phi_1(T) - \phi_2(T))^2\rangle)$. 
The connected correlations of the noise can then be related to $\langle \phi_1(T) \phi_2(T)\rangle - \langle \phi_1(T) \rangle \langle \phi_2(T) \rangle$. Assuming the distribution function is stationary in time, this phase correlation function measures the integrated two-time correlator 
\begin{equation}\label{phasecorr}
C_\phi = 2 \gamma^2 T \int_0^T \mathrm{d}t \; \left\langle (\hat n_1 \cdot \vec{B}(\vec\rho_1, t))(\hat n_2 \cdot \vec{B}(\vec{\rho}_2, 0)) \right\rangle.
\end{equation}
Often, dynamical decoupling~\cite{barry2020,machado2022} is employed to effectively filter out undesired frequency components such as the DC contribution. However, in the frequency range relevant to the NV center---around MHz to GHz---the noise spectrum of the metal is essentially white. As a result, dynamical decoupling would not affect the phase accumulation and we can simply integrate the magnetic noise.

\begin{figure}[tb]
\center
\includegraphics[width=0.55\linewidth]{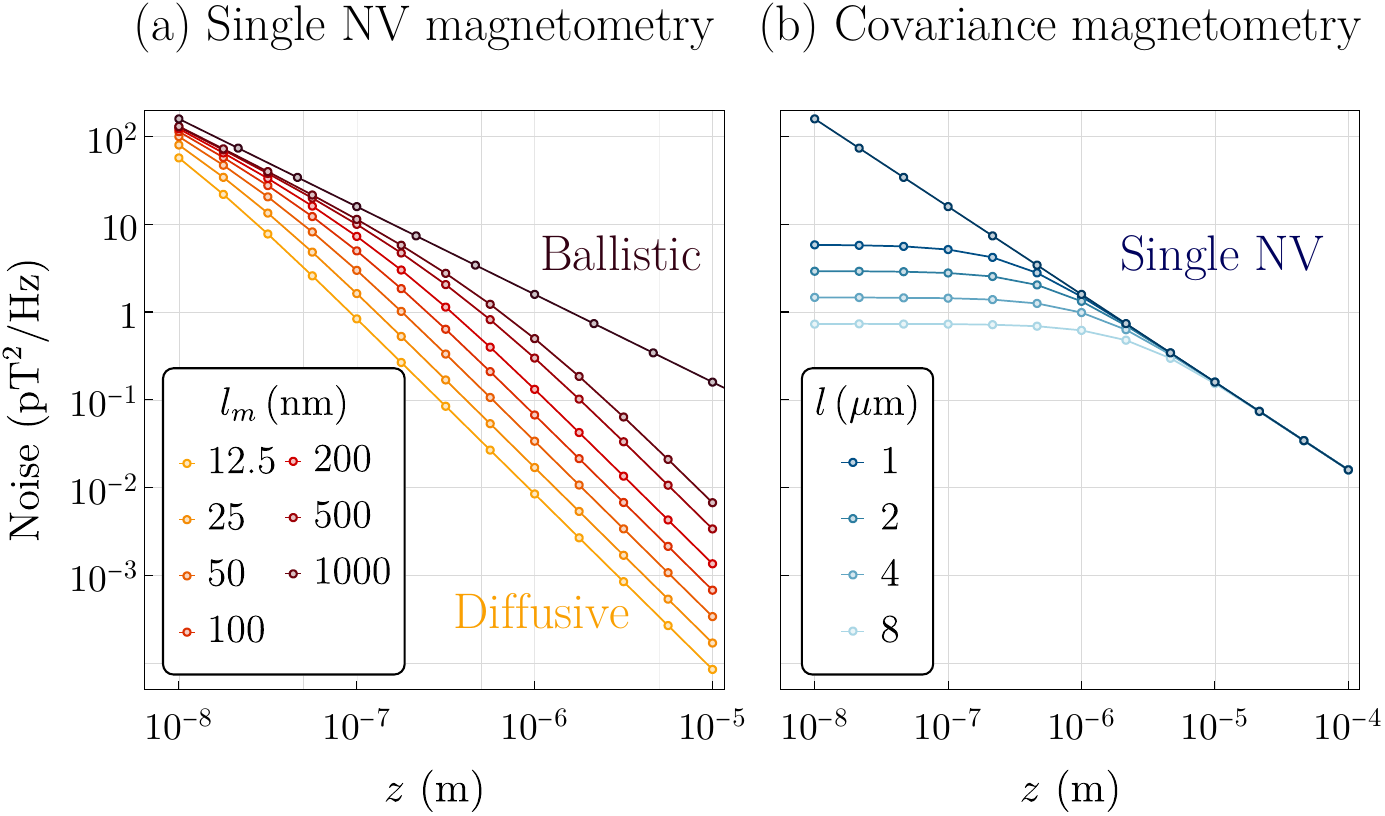}
\caption{\label{fig:experiment}(a) Scaling of noise with the depth $z$ in single-NV magnetometry for mean free paths $l_m$ ranging from $12.5$ to $1000$~nm, as well as for perfectly ballistic transport.  (b) Scaling of noise with depth $z$ in covariance magnetometry for NV separations $l = 1,2,4,8$ $\mu$m and under ballistic transport. The scaling obtained with single-NV magnetometry is also shown for reference. Note the key distinction that bringing the NV centers closer to the sample, beyond a certain threshold, does not yield additional sensitivity in covariance magnetometry experiments.}
\end{figure}

\section{Equilibrium fluctuations}

We begin by applying this formalism to estimate the equilibrium fluctuations in a two-dimensional metal. Since we expect correlations to be strongest in the ballistic regime, we focus on graphene, which has a long mean free path. In Fig.~\ref{fig:experiment}(a), we plot the noise power sensed by a single NV center as a function of its depth for scattering lengths varying from 12.5 nm to 1 $\mu$m. We set the chemical potential to $0.2$ eV (which is achievable for graphene by gate-tuning~\cite{Andersen2019a}) and the temperature to 80 K. The brown line shows the ideal noise power when the transport is ballistic, scaling as $1/z$ (see below). In the presence of scattering, the scaling of the noise transitions to  $1/z^2$ when $z$ is bigger than the scattering length; this prediction is consistent with previous calculations~\cite{Kolkowitz2015a, PhysRevB.95.155107}. The strength of the noise can be calculated to scale linearly with temperature in the range considered here (this follows from the fluctuation-dissipation theorem since the resistivity is not strongly temperature dependent).

We turn next to covariance magnetometry in Fig.~\ref{fig:experiment}(b), using the same parameters as in panel~(a), but focusing on purely ballistic transport (so the two NVs are taken to be within a mean free path of each other).
The signal strength now depends on both the height $z$ of the NVs and the separation $l$ between them. In stark contrast to the single-NV case, the signal strength cannot be enhanced by bringing the NVs arbitrarily close to the sample. 
We provide a simple dimensional argument for this. Note that in two dimensions, the factors of $\vec{r}$ in the numerator and denominator of Eq.~\eqref{biot_savart} cancel. 
Thus, the only explicit spatial dependence comes from the correlation function $f_2$, which (in a ballistic system) takes the form
\begin{equation}
f_2 (\vec{r}_1, t; \vec{r}_2, 0) \sim \int d^2\vec{k} \, \delta^{(2)}\left(\vec{r}_1 - \vec{r}_2 - \vec{v}_{\vec k} t\right),
\end{equation}
i.e., it consists of independent contributions from fluctuations at each $\vec{k}$ mode that propagate ballistically from $\vec{r}_1$ to $\vec{r}_2$ in time $t$. The notation $\delta^{(2)}$ denotes a two-dimensional $\delta$-function.
The total phase accumulated~\eqref{phasecorr} is integrated over time, so the time-integrated correlator is a one-dimensional $\delta$-function, forcing $\vec{r}_1$ and $\vec{r}_2$ to lie on a straight line with an orientation set by $\vec k$. Thus, for each mode $\vec{k}$, this constraint contributes a factor of $1/z$ upon rescaling.

There are two asymptotic regimes of behavior, set by the ratio $z/l$. When $z$\,$\gg$\,$l$, the two NVs are essentially at the same point and see the same quasiparticle trajectories. In this limit, the two-NV result reduces to the familiar single-NV result for ballistic systems~\cite{Kolkowitz2015a}: since the phase space of quasiparticles is independent of $z$, the integrated correlator~\eqref{phasecorr} scales as $1/z$. However, when $z$\,$\ll$\,$l$, the only contribution to the covariance comes from quasiparticles that pass through two discs of radius $\sim z$, one centered underneath each NV (Fig.~\ref{fig:setup}). The momenta of these quasiparticles come from an angular range $z/l$, and this factor of $z$ cancels out the overall $1/z$ noise strength per mode. To summarize, the increased noise strength from moving the NVs closer to the sample is canceled out by the more restricted phase space from requiring quasiparticles to pass under both NVs. 

Although we derived it in the specific context of the collisionless Boltzmann equation, the extra factor of $z/l$ is general: in any genuinely two-dimensional geometry (as opposed to e.g. decoupled chains), moving the NV farther from the sample compromises noise strength but gains phase space. In the ballistic case, this tradeoff completely eliminates the advantage of moving NVs near the sample (and might in fact lead to worse performance once one accounts for surface effects); in more general contexts, it might only decrease the advantage. 

\begin{figure*}[tb]
\includegraphics[width=\linewidth]{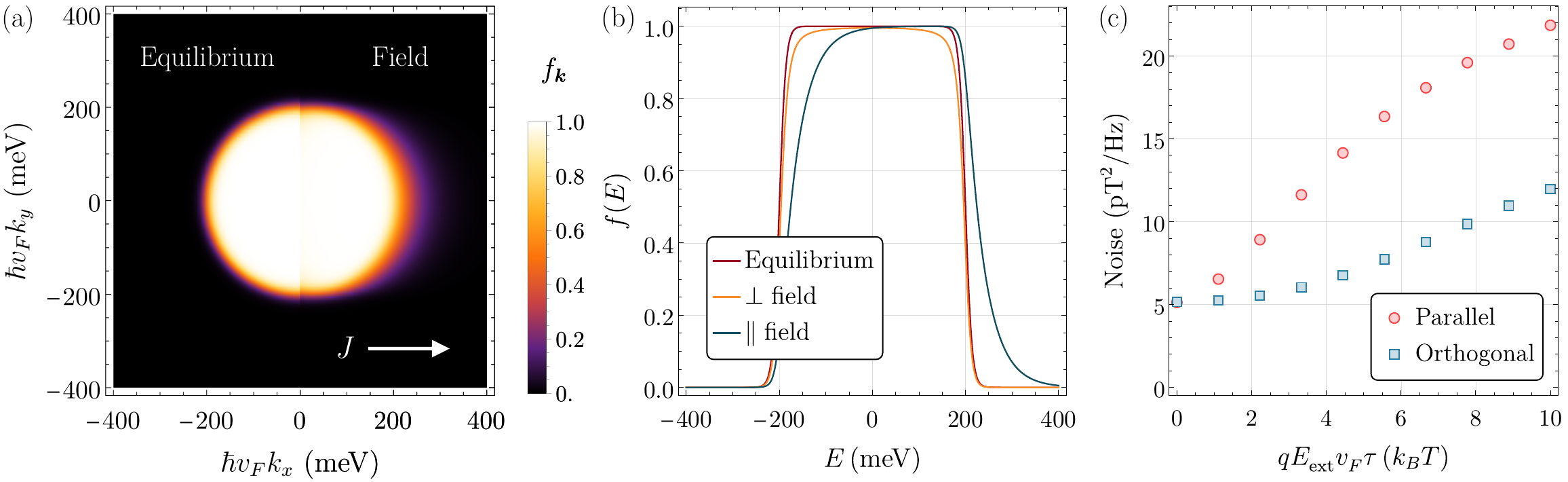}
\caption{\label{fig:nonequilibrium}(a) Distribution function in thermal equilibrium (left half) and under external field (right half). (b) Cross-section of the distribution function in thermal equilibrium (red) and under external field perpendicular (orange) or parallel (blue) to the external drive. (c) Magnetic noise seen by the covariance magnetometry as a function of the current density when the axis connecting the NVs is aligned with the current (red) or orthogonal to it (blue). The increase in the orthogonal noise power is because $q E_{\mathrm{ext}} v_F \tau$ approaches the chemical potential near $10 \, k_B T$, so the Fermi-liquid picture starts to break down.}
\end{figure*}

\section{Nonequilibrium fluctuations}

So far, we have focused on equilibrium noise, but our formalism extends naturally to the nonequilibrium case. We now discuss a simple nonequilibrium setting, in which a uniform current is applied to the sample, and compute the spatial structure of current noise in that setting. 
We will argue that the angular dependence of the nonequilibrium noise carries significant information about the mechanisms behind both transport and momentum/energy relaxation.

First, we explicitly solve for the nonequilibrium steady-state distribution function $f(\vec{r},\vec{k},t)$ and the noise correlator $f_2$ in the relaxation-time approximation. The applied electric field distorts the Fermi sea as shown in Fig.~\ref{fig:nonequilibrium}(a,b). This distorted nonequilibrium distribution function is far from zero or one in a window of energy width $q E_{\mathrm{ext}} v_F \tau$, where $E_{\mathrm{ext}}$ is the applied field strength, $v_F$ is the Fermi velocity, and $\tau$ is the scattering lifetime. This width can be understood on dimensional grounds: the electric field imparts momentum to the system at a rate $qE_{\mathrm{ext}}$, and balancing this against the relaxation rate $\tau^{-1}$ gives a steady-state momentum $q E_{\mathrm{ext}} \tau$, which matches the energy width above.
It is helpful to define an angle-dependent effective ``temperature'' $T_{\mathrm{eff}} = q E_{\mathrm{ext}} v_F \tau/k_B$ for the nonequilibrium fluctuations. In the case of graphene, at the chemical potential of 0.2 eV and assuming a mean free path of 1\,$\mu$m, a current density of 1.6 mA/$\mu$m corresponds to an effective temperature of 800 K, so one would expect a roughly tenfold enhancement in noise strength relative to equilibrium.

Driving a current also heats up the sample, so how can one tell if the enhanced fluctuations are due to heating or due to nonequilibrium effects? The \emph{anisotropic} enhancement of noise gives a useful signature of the nonequilibrium nature of the distribution, which can be probed using covariance magnetometry. If the two NV centers are aligned with the current flow, they should observe enhanced fluctuations. On the other hand, NV centers aligned perpendicular to the current flow should not see enhanced fluctuations. This argument is numerically confirmed in Fig.~\ref{fig:nonequilibrium}(c). Since heating results in isotropically enhanced fluctuations, this protocol can verify the genuine nonequilibrium nature of the fluctuations. We also emphasize that this anisotropic noise enhancement cannot be seen with a single NV. Crucially, the calculation above relies on the momentum anisotropy of the distribution function. Thus, while it is common in the literature to integrate out the momentum \cite{nagaev1995influence,kozub1995shot}, we cannot do so here because the momentum dependence is central to finding the nonequilibrium noise.

By using RTA, we implicitly assumed that the processes which relax energy and momentum are the same. It would be interesting to explore the effects of an electric field for more general collision integrals, e.g., using the methods in Ref.~\onlinecite{PhysRevB.108.L121401}. At the level of dimensional analysis, we can estimate the effects when energy and momentum relax at separate rates, $\tau_E$ and $\tau_P$, respectively. The steady-state energy width of the distribution function is then $q E_{\mathrm{ext}} v_F \tau_E$ while the net momentum is $q E_{\mathrm{ext}} \tau_P$. 
When $\tau_E = \tau_P$, displacing the distribution parallel to the current is sufficient to generate the net steady-state momentum, but for $\tau_E > \tau_P$ (e.g., when either impurity scattering or electron-electron interactions are dominant), the net momentum relaxes faster than the energy, so part of the distribution displaces perpendicular to the current. Hence, the noise anisotropy decreases when $\tau_E > \tau_P$, and dimensional analysis yields the ratio between the perpendicular and parallel effective temperatures to be $\sim 1-\tau_P/\tau_E$. We therefore expect covariance magnetometry to distinguish the relaxation channels by measuring the noise anisotropy.%

\section{Discussion and outlook}

In this work, we have computed the spatiotemporal correlations of nonequilibrium current noise in two-dimensional metals. Within the Boltzmann framework (augmented to include noise terms), we quantitatively estimate the detectability of nonequilibrium noise correlations in present-day experiments using NV centers placed above a metallic sample~\cite{Rovny2022a}. We argued that the spatial structure of nonequilibrium noise---in particular, its spatial anisotropy---carries information about the nature of momentum and energy relaxation in Fermi liquids. 
We emphasize that covariance magnetometry probes \emph{local} nonequilibrium noise. This is in sharp contrast to global transport experiments that probe global noise averaged across the sample, rendering the results system-size-dependent \cite{blanter2000shot,tikhonov2020spatial,wienand2023emergence,rosenberg2023dynamics}. Local noise also directly reveals the anisotropy and the correlation length, providing additional information about the system.
Although explicit calculations outside this Fermi-liquid regime are much more challenging, the physical mechanism causing the anisotropy clearly also applies to more exotic systems such as non-Fermi liquids with a well-defined Fermi surface. It would be interesting to adapt recent field theories of nonequilibrium matter~\cite{lucas2023} to explore covariance magnetometry in these systems.

It would be particularly interesting to extend our results to the case of two-dimensional superconductors, or anomalous metals that are near a superconducting instability~\cite{zhang2026detecting}. In these systems, electrical resistance is believed to be due to vortex motion. The current stochastically dissociates loosely bound vortex pairs, and these drift perpendicular to the applied current. Thus, two NVs whose separation is perpendicular to the current flow would see temporal correlations due to the creation and flow of vortices between them. The arguments we have outlined above suggest that current noise should therefore be enhanced \emph{perpendicular} to the current flow, but not parallel to it, in contrast to the behavior of metals with electron-like quasiparticles. Observing this anisotropy would be a stringent test of the theory of anomalous metals~\cite{RevModPhys.91.011002}. 

Lastly, NV magnetometry can also perform frequency-resolved measurements through dynamical decoupling. The accessible frequency window is typically between MHz and GHz. In metals, all characteristic dynamical timescales are much shorter than this window, so the NV center effectively experiences white noise with no resolvable spectral features. It would be interesting to study systems with slow degrees of freedom (such as vortices), whose fluctuations may produce spectral structure within the NV detection window. At the opposite extreme, accessing frequencies well above $\sim$\,1\,GHz, where the noise spectra of metals begin to roll off due to electronic relaxation rates, requires sensing protocols beyond standard dynamical decoupling, such as $T_1$ relaxometry near the NV ground-state splitting~\cite{Andersen2019a,10.1103/physrevlett.129.087701} or off-resonant driving schemes. These higher-frequency probes could in principle distinguish microscopic relaxation channels in metals or correlated phases, and represent a natural extension of the covariance-magnetometry protocols discussed here.

\acknowledgments{We thank Nathalie de Leon, Eugene Demler, Andrew Lucas, Jamir Marino, and Jared Rovny for helpful discussions and collaborations on related topics. S.G. and Y.Z. acknowledge support from NSF QuSEC-TAQS OSI 2326767. R.S. is supported by the Princeton Quantum Initiative Fellowship.}

\appendix

\newpage
\newpage
\begin{widetext}

\section{}
\label{sec:app}

In this appendix, we first present a solution of the Boltzmann equation for a collision integral that is local in momentum space, corresponding to scattering off weak or long-ranged impurities. Second, we present a more detailed derivation of the scaling of the equilibrium noise with the height of the NV centers above the surface. 

\subsection{Scattering off smooth potentials}

When scattering is local in momentum space, it leads to momentum diffusion. In this case, we can analytically keep track of the collision functional to find the nonequal-time correlations. To begin, we expand $f$ around the steady state $f_{\mathrm{ss}}$ by defining $\delta f = f - f_{\mathrm{ss}}$. To first order,
\begin{align}
    \partial_t \delta f &=-(\vec{v}_{\vec k} \cdot \nabla_r + \vec{F}_{\mathrm{ext}} \cdot \nabla_k) \delta f - I_{\mathrm{col}}[\delta f], \\
     I_{\mathrm{col}}[\delta f] &=\sum_{\Delta \vec{k}}W_{\Delta \vec{k}}\left(\delta f(\vec{r},\vec{k},t)-\delta f(\vec{r},\vec{k}+\Delta\vec{k},t)\right).
\end{align}
In this expansion, we have assumed time-reversal symmetry in the scattering such that $W_{\Delta \vec{k}}=W_{-\Delta \vec{k}}$. To proceed, we rewrite the collision functional in the integral form
\begin{equation}
    I_{\mathrm{col}}[\delta f] = \int d \Delta \vec{k} \; W_{\Delta \vec{k}}\left(\delta f(\vec{r},\vec{k},t)-\delta f(\vec{r},\vec{k}+\Delta\vec{k},t)\right).
\end{equation}
By virtue of the locality of the scattering in momentum space, we can Taylor-expand the scattering rate $W_{\Delta \vec{k}}$ and $\delta f(\vec{r},\vec{k}+\Delta\vec{k},t)$ in $\Delta \vec{k}$:
\begin{align}
    W_{\Delta \vec{k}} &= W^{(0)} + \frac{1}{2}W^{(2)}_{ab} \Delta k_a \Delta k_b + \mathcal{O}(\Delta k^4), \\
    \delta f(\vec{r},\vec{k}+\Delta\vec{k},t)-\delta f(\vec{r},\vec{k},t) &= \frac{\partial \delta f}{\partial k_a} \Delta k_a + \frac{1}{2}\frac{\partial^2 \delta f}{\partial k_a\partial k_b} \Delta k_a \Delta k_b + \mathcal{O}(\Delta k^3).
\end{align}
Note that the term first-order in $\Delta\vec{k}$ vanishes because of time-reversal symmetry. We will discuss the case when time-reversal symmetry is broken, and the consequences thereof, in future work. Only the even-order term in the integral survives, so to leading order,
\begin{align}
    I_{\mathrm{col}}[\delta f] &= -\int d \Delta \vec{k}\, \frac{1}{2} W^{(0)}\Delta k_a \Delta k_b \frac{\partial^2 \delta f}{\partial k_a\partial k_b} = -\delta_{ab} \mathcal{B} \frac{\partial^2 \delta f}{\partial k_a\partial k_b} = -\mathcal{B} \nabla^2_{\vec{k}} \delta f,
\end{align}
where we define $\mathcal{B}$ as the value of the integral when $a=b$ (the overall minus sign arises because the integrand involves $\delta f(\vec k) - \delta f(\vec k+\Delta\vec k)$, opposite in sign to the difference Taylor-expanded above). We therefore eventually arrive at the following linearized Boltzmann equation:
\begin{equation}
     \partial_t \delta f = -\left(\vec{v}_{\vec k} \cdot \nabla_r + \vec{F}_{\mathrm{ext}} \cdot \nabla_k - \mathcal{B} \nabla^2_{\vec{k}} \right) \delta f.
\end{equation}
This resembles the Fokker-Planck equation of a Brownian particle with the equations of motion
\begin{align}
    \partial_t \vec{r} &= \frac{\vec{k}}{m}, \\
    \partial_t \vec{k} &= \vec{F}_{\mathrm{ext}} -\frac{\gamma}{m} \vec{k} + \vec{\xi},
\end{align}
where the white noise $\vec{\xi}$ is connected to the dissipation coefficient $\gamma$ via the fluctuation-dissipation theorem: $\langle \xi_a(t) \xi_b(t')\rangle = 2 \gamma k_B T\, \delta_{ab}\, \delta (t-t')$. Nevertheless, we do not and should not assume any thermodynamic relations since we want our formalism to be applicable in the nonequilibrium setup as well. Indeed, based on our second insight, the dynamics of $f_2$ are governed by the same Boltzmann equation subject to the initial condition Eq.~\eqref{equaltime},
\begin{align} \label{local_scattering_f2}
\begin{split}
&\partial_{\Delta t}f_2(\vec{r},\vec{k},t+\Delta t;\vec{r'},\vec{k'},t)= -\left(\vec{v}_{\vec k} \cdot \nabla_r + \vec{F}_{\mathrm{ext}} \cdot \nabla_k - \mathcal{B} \nabla^2_{\vec{k}} \right)f_2.
\end{split}
\end{align}
Initially, $f_2$ is a delta function at $(\vec{r}',\vec{k'})$. As $\Delta t$ increases, $f_2$ starts to spread out, undergoing drift-diffusion in phase space. 

\subsection{Scaling of noise with NV height}

We now provide a more detailed derivation of the scaling of the noise when $z\ll l$ as observed in Fig.~\ref{fig:experiment}(b). We begin by factoring $z$ out of Eq.~(\ref{biot_savart}) as
 \begin{equation}
     B_y(\vec{0},t)=-\frac{\mu_0}{4 \pi}\frac{1}{z^2}\int d^2\vec{r} \frac{J_x(\vec{r},t) }{(|\vec{r}/z|^2+1)^{3/2}}.
 \end{equation}
 Assuming a uniform current distribution, the integral can be cut off at $r/z$ being some $\mathcal{O}(1)$ value, so we will truncate the integral to a disk of radius $R \sim z$. Additionally, $(|\vec{r}/z|^2+1)^{3/2}$ is an $\mathcal{O}(1)$ number inside the disk, so we drop it for the purposes of our scaling analysis. Next, we evaluate the magnetic noise with this truncation as
 \begin{equation}
     \langle B_y(\vec{0},t)B_y(\vec{l},t')\rangle \propto \frac{1}{z^4}\int_{\text{Disk 1}} d^2\vec{r} \int_{\text{Disk 2}}d^2\vec{r'}\, \langle J_x(\vec{r},t)J_x(\vec{r'},t')\rangle,
 \end{equation}
 where ``Disk 1'' and ``Disk 2'' are the truncated disks at the two NV centers separated by $l \gg R$. Roughly speaking, the correlations between every set of $\vec{r},\vec{r}'$ pairs contribute a factor of $R^4$ that cancels out the $1/z^4$. This explains the saturation of the noise at $z \ll l $. To be concrete, we expand $\langle J_x(\vec{r},t)J_x(\vec{r'},t')\rangle$ in polar coordinates
 \begin{align}
     \langle J_x(\vec{r},t)J_x(\vec{r'},t')\rangle \propto \int d\theta \, k \, dk\, k^2 \cos^2(\theta) f_{\mathrm{ss}}(k)(1-f_{\mathrm{ss}}(k)) \delta(\vec{r'}-\vec{r}-\vec{v}_{\vec{k}}(t'-t))\label{current_corr_scaling};
 \end{align}
we integrate over the polar coordinates with $\theta=0$ being the direction along which the NVs are aligned. Given that the two disks are far away, only small angles $\theta$  between $\pm R/l$ contribute to the integral, so we can treat $\cos^2(\theta)$ as an $\mathcal{O}(1)$ number. Lastly, the phase noise involves integrating the magnetic noise over $t$ and $t'-t$
 \begin{equation}
     \langle \phi_1 \phi_2 \rangle \propto \int dt \, d(t'-t) \langle B_y(\vec{0},t)B_y(\vec{l},t')\rangle.
 \end{equation}
 We first integrate over $\vec{r}'$ to remove the delta function in Eq.~\eqref{current_corr_scaling}. This constrains the integral of $\theta$ to be between $\pm R/l$ and the integral of $t'-t$ to be between $(l \pm R)/v_F$. Together with integrating $\vec{r}$, this results in a net $R^4$ contribution, canceling out the $1/z^4$ factor, so the noise is independent of the depth when $l \gg R$.

 In the opposite regime where $l \ll R$, the small-angle condition of $\theta$ lying between $\pm R/l$ no longer holds. Therefore, the scaling analysis breaks down and one recovers the single-NV scaling behavior as seen in Fig.~\ref{fig:experiment}(b).

\end{widetext}


\bibliography{nvnoise}

\providecommand{\noopsort}[1]{}\providecommand{\singleletter}[1]{#1}%
\begin{thebibliography}{54}%
\makeatletter
\providecommand \@ifxundefined [1]{%
 \@ifx{#1\undefined}
}%
\providecommand \@ifnum [1]{%
 \ifnum #1\expandafter \@firstoftwo
 \else \expandafter \@secondoftwo
 \fi
}%
\providecommand \@ifx [1]{%
 \ifx #1\expandafter \@firstoftwo
 \else \expandafter \@secondoftwo
 \fi
}%
\providecommand \natexlab [1]{#1}%
\providecommand \enquote  [1]{``#1''}%
\providecommand \bibnamefont  [1]{#1}%
\providecommand \bibfnamefont [1]{#1}%
\providecommand \citenamefont [1]{#1}%
\providecommand \href@noop [0]{\@secondoftwo}%
\providecommand \href [0]{\begingroup \@sanitize@url \@href}%
\providecommand \@href[1]{\@@startlink{#1}\@@href}%
\providecommand \@@href[1]{\endgroup#1\@@endlink}%
\providecommand \@sanitize@url [0]{\catcode `\\12\catcode `\$12\catcode
  `\&12\catcode `\#12\catcode `\^12\catcode `\_12\catcode `\%12\relax}%
\providecommand \@@startlink[1]{}%
\providecommand \@@endlink[0]{}%
\providecommand \url  [0]{\begingroup\@sanitize@url \@url }%
\providecommand \@url [1]{\endgroup\@href {#1}{\urlprefix }}%
\providecommand \urlprefix  [0]{URL }%
\providecommand \Eprint [0]{\href }%
\providecommand \doibase [0]{https://doi.org/}%
\providecommand \selectlanguage [0]{\@gobble}%
\providecommand \bibinfo  [0]{\@secondoftwo}%
\providecommand \bibfield  [0]{\@secondoftwo}%
\providecommand \translation [1]{[#1]}%
\providecommand \BibitemOpen [0]{}%
\providecommand \bibitemStop [0]{}%
\providecommand \bibitemNoStop [0]{.\EOS\space}%
\providecommand \EOS [0]{\spacefactor3000\relax}%
\providecommand \BibitemShut  [1]{\csname bibitem#1\endcsname}%
\let\auto@bib@innerbib\@empty
\bibitem [{\citenamefont {Qu}\ \emph {et~al.}(2019)\citenamefont {Qu},
  \citenamefont {Benz}, \citenamefont {Rogalla}, \citenamefont {Tew},
  \citenamefont {White},\ and\ \citenamefont {Zhou}}]{qu2019johnson}%
  \BibitemOpen
  \bibfield  {author} {\bibinfo {author} {\bibfnamefont {J.~F.}\ \bibnamefont
  {Qu}}, \bibinfo {author} {\bibfnamefont {S.~P.}\ \bibnamefont {Benz}},
  \bibinfo {author} {\bibfnamefont {H.}~\bibnamefont {Rogalla}}, \bibinfo
  {author} {\bibfnamefont {W.}~\bibnamefont {Tew}}, \bibinfo {author}
  {\bibfnamefont {D.~R.}\ \bibnamefont {White}},\ and\ \bibinfo {author}
  {\bibfnamefont {K.~L.}\ \bibnamefont {Zhou}},\ }\bibfield  {title} {\bibinfo
  {title} {Johnson noise thermometry},\ }\href
  {https://doi.org/10.1088/1361-6501/ab3526} {\bibfield  {journal} {\bibinfo
  {journal} {Meas. Sci. Technol.}\ }\textbf {\bibinfo {volume} {30}},\ \bibinfo
  {pages} {112001} (\bibinfo {year} {2019})}\BibitemShut {NoStop}%
\bibitem [{\citenamefont {Kolkowitz}\ \emph {et~al.}(2015)\citenamefont
  {Kolkowitz}, \citenamefont {Safira}, \citenamefont {High}, \citenamefont
  {Devlin}, \citenamefont {Choi}, \citenamefont {Unterreithmeier},
  \citenamefont {Patterson}, \citenamefont {Zibrov}, \citenamefont
  {Manucharyan}, \citenamefont {Park},\ and\ \citenamefont
  {Lukin}}]{Kolkowitz2015a}%
  \BibitemOpen
  \bibfield  {author} {\bibinfo {author} {\bibfnamefont {S.}~\bibnamefont
  {Kolkowitz}}, \bibinfo {author} {\bibfnamefont {A.}~\bibnamefont {Safira}},
  \bibinfo {author} {\bibfnamefont {A.~A.}\ \bibnamefont {High}}, \bibinfo
  {author} {\bibfnamefont {R.~C.}\ \bibnamefont {Devlin}}, \bibinfo {author}
  {\bibfnamefont {S.}~\bibnamefont {Choi}}, \bibinfo {author} {\bibfnamefont
  {Q.~P.}\ \bibnamefont {Unterreithmeier}}, \bibinfo {author} {\bibfnamefont
  {D.}~\bibnamefont {Patterson}}, \bibinfo {author} {\bibfnamefont {A.~S.}\
  \bibnamefont {Zibrov}}, \bibinfo {author} {\bibfnamefont {V.~E.}\
  \bibnamefont {Manucharyan}}, \bibinfo {author} {\bibfnamefont
  {H.}~\bibnamefont {Park}},\ and\ \bibinfo {author} {\bibfnamefont {M.~D.}\
  \bibnamefont {Lukin}},\ }\bibfield  {title} {\bibinfo {title} {Probing
  {Johnson} noise and ballistic transport in normal metals with a single-spin
  qubit},\ }\href {https://doi.org/10.1126/science.aaa4298} {\bibfield
  {journal} {\bibinfo  {journal} {Science}\ }\textbf {\bibinfo {volume}
  {347}},\ \bibinfo {pages} {1129} (\bibinfo {year} {2015})}\BibitemShut
  {NoStop}%
\bibitem [{\citenamefont {Agarwal}\ \emph {et~al.}(2017)\citenamefont
  {Agarwal}, \citenamefont {Schmidt}, \citenamefont {Halperin}, \citenamefont
  {Oganesyan}, \citenamefont {Zar\'and}, \citenamefont {Lukin},\ and\
  \citenamefont {Demler}}]{PhysRevB.95.155107}%
  \BibitemOpen
  \bibfield  {author} {\bibinfo {author} {\bibfnamefont {K.}~\bibnamefont
  {Agarwal}}, \bibinfo {author} {\bibfnamefont {R.}~\bibnamefont {Schmidt}},
  \bibinfo {author} {\bibfnamefont {B.}~\bibnamefont {Halperin}}, \bibinfo
  {author} {\bibfnamefont {V.}~\bibnamefont {Oganesyan}}, \bibinfo {author}
  {\bibfnamefont {G.}~\bibnamefont {Zar\'and}}, \bibinfo {author}
  {\bibfnamefont {M.~D.}\ \bibnamefont {Lukin}},\ and\ \bibinfo {author}
  {\bibfnamefont {E.}~\bibnamefont {Demler}},\ }\bibfield  {title} {\bibinfo
  {title} {Magnetic noise spectroscopy as a probe of local electronic
  correlations in two-dimensional systems},\ }\href
  {https://doi.org/10.1103/PhysRevB.95.155107} {\bibfield  {journal} {\bibinfo
  {journal} {Phys. Rev. B}\ }\textbf {\bibinfo {volume} {95}},\ \bibinfo
  {pages} {155107} (\bibinfo {year} {2017})}\BibitemShut {NoStop}%
\bibitem [{\citenamefont {de~Picciotto}\ \emph {et~al.}(1998)\citenamefont
  {de~Picciotto}, \citenamefont {Reznikov}, \citenamefont {Heiblum},
  \citenamefont {Umansky}, \citenamefont {Bunin},\ and\ \citenamefont
  {Mahalu}}]{de1998direct}%
  \BibitemOpen
  \bibfield  {author} {\bibinfo {author} {\bibfnamefont {R.}~\bibnamefont
  {de~Picciotto}}, \bibinfo {author} {\bibfnamefont {M.}~\bibnamefont
  {Reznikov}}, \bibinfo {author} {\bibfnamefont {M.}~\bibnamefont {Heiblum}},
  \bibinfo {author} {\bibfnamefont {V.}~\bibnamefont {Umansky}}, \bibinfo
  {author} {\bibfnamefont {G.}~\bibnamefont {Bunin}},\ and\ \bibinfo {author}
  {\bibfnamefont {D.}~\bibnamefont {Mahalu}},\ }\bibfield  {title} {\bibinfo
  {title} {Direct observation of a fractional charge},\ }\href
  {https://doi.org/10.1016/S0921-4526(98)00139-2} {\bibfield  {journal}
  {\bibinfo  {journal} {Physica B}\ }\textbf {\bibinfo {volume} {249}},\
  \bibinfo {pages} {395} (\bibinfo {year} {1998})}\BibitemShut {NoStop}%
\bibitem [{\citenamefont {Reznikov}\ \emph {et~al.}(1999)\citenamefont
  {Reznikov}, \citenamefont {de~Picciotto}, \citenamefont {Griffiths},
  \citenamefont {Heiblum},\ and\ \citenamefont
  {Umansky}}]{reznikov1999observation}%
  \BibitemOpen
  \bibfield  {author} {\bibinfo {author} {\bibfnamefont {M.}~\bibnamefont
  {Reznikov}}, \bibinfo {author} {\bibfnamefont {R.}~\bibnamefont
  {de~Picciotto}}, \bibinfo {author} {\bibfnamefont {T.~G.}\ \bibnamefont
  {Griffiths}}, \bibinfo {author} {\bibfnamefont {M.}~\bibnamefont {Heiblum}},\
  and\ \bibinfo {author} {\bibfnamefont {V.}~\bibnamefont {Umansky}},\
  }\bibfield  {title} {\bibinfo {title} {Observation of quasiparticles with
  one-fifth of an electron's charge},\ }\href {https://doi.org/10.1038/20384}
  {\bibfield  {journal} {\bibinfo  {journal} {Nature}\ }\textbf {\bibinfo
  {volume} {399}},\ \bibinfo {pages} {238} (\bibinfo {year}
  {1999})}\BibitemShut {NoStop}%
\bibitem [{\citenamefont {Kane}\ and\ \citenamefont
  {Fisher}(1994)}]{PhysRevLett.72.724}%
  \BibitemOpen
  \bibfield  {author} {\bibinfo {author} {\bibfnamefont {C.~L.}\ \bibnamefont
  {Kane}}\ and\ \bibinfo {author} {\bibfnamefont {M.~P.~A.}\ \bibnamefont
  {Fisher}},\ }\bibfield  {title} {\bibinfo {title} {{Nonequilibrium noise and
  fractional charge in the quantum Hall effect}},\ }\href
  {https://doi.org/10.1103/PhysRevLett.72.724} {\bibfield  {journal} {\bibinfo
  {journal} {Phys. Rev. Lett.}\ }\textbf {\bibinfo {volume} {72}},\ \bibinfo
  {pages} {724} (\bibinfo {year} {1994})}\BibitemShut {NoStop}%
\bibitem [{\citenamefont {Chamon}\ \emph {et~al.}(1995)\citenamefont {Chamon},
  \citenamefont {Freed},\ and\ \citenamefont {Wen}}]{PhysRevB.51.2363}%
  \BibitemOpen
  \bibfield  {author} {\bibinfo {author} {\bibfnamefont {C.~d.~C.}\
  \bibnamefont {Chamon}}, \bibinfo {author} {\bibfnamefont {D.~E.}\
  \bibnamefont {Freed}},\ and\ \bibinfo {author} {\bibfnamefont {X.~G.}\
  \bibnamefont {Wen}},\ }\bibfield  {title} {\bibinfo {title} {{Tunneling and
  quantum noise in one-dimensional Luttinger liquids}},\ }\href
  {https://doi.org/10.1103/PhysRevB.51.2363} {\bibfield  {journal} {\bibinfo
  {journal} {Phys. Rev. B}\ }\textbf {\bibinfo {volume} {51}},\ \bibinfo
  {pages} {2363} (\bibinfo {year} {1995})}\BibitemShut {NoStop}%
\bibitem [{\citenamefont {Fendley}\ \emph {et~al.}(1995)\citenamefont
  {Fendley}, \citenamefont {Ludwig},\ and\ \citenamefont
  {Saleur}}]{PhysRevLett.75.2196}%
  \BibitemOpen
  \bibfield  {author} {\bibinfo {author} {\bibfnamefont {P.}~\bibnamefont
  {Fendley}}, \bibinfo {author} {\bibfnamefont {A.~W.~W.}\ \bibnamefont
  {Ludwig}},\ and\ \bibinfo {author} {\bibfnamefont {H.}~\bibnamefont
  {Saleur}},\ }\bibfield  {title} {\bibinfo {title} {{Exact Nonequilibrium dc
  Shot Noise in Luttinger Liquids and Fractional Quantum Hall Devices}},\
  }\href {https://doi.org/10.1103/PhysRevLett.75.2196} {\bibfield  {journal}
  {\bibinfo  {journal} {Phys. Rev. Lett.}\ }\textbf {\bibinfo {volume} {75}},\
  \bibinfo {pages} {2196} (\bibinfo {year} {1995})}\BibitemShut {NoStop}%
\bibitem [{\citenamefont {Feldman}\ and\ \citenamefont
  {Heiblum}(2017)}]{PhysRevB.95.115308}%
  \BibitemOpen
  \bibfield  {author} {\bibinfo {author} {\bibfnamefont {D.~E.}\ \bibnamefont
  {Feldman}}\ and\ \bibinfo {author} {\bibfnamefont {M.}~\bibnamefont
  {Heiblum}},\ }\bibfield  {title} {\bibinfo {title} {Why a noninteracting
  model works for shot noise in fractional charge experiments},\ }\href
  {https://doi.org/10.1103/PhysRevB.95.115308} {\bibfield  {journal} {\bibinfo
  {journal} {Phys. Rev. B}\ }\textbf {\bibinfo {volume} {95}},\ \bibinfo
  {pages} {115308} (\bibinfo {year} {2017})}\BibitemShut {NoStop}%
\bibitem [{\citenamefont {Wang}\ \emph {et~al.}(2024)\citenamefont {Wang},
  \citenamefont {Setty}, \citenamefont {Sur}, \citenamefont {Chen},
  \citenamefont {Paschen}, \citenamefont {Natelson},\ and\ \citenamefont
  {Si}}]{wang2022shot}%
  \BibitemOpen
  \bibfield  {author} {\bibinfo {author} {\bibfnamefont {Y.}~\bibnamefont
  {Wang}}, \bibinfo {author} {\bibfnamefont {C.}~\bibnamefont {Setty}},
  \bibinfo {author} {\bibfnamefont {S.}~\bibnamefont {Sur}}, \bibinfo {author}
  {\bibfnamefont {L.}~\bibnamefont {Chen}}, \bibinfo {author} {\bibfnamefont
  {S.}~\bibnamefont {Paschen}}, \bibinfo {author} {\bibfnamefont
  {D.}~\bibnamefont {Natelson}},\ and\ \bibinfo {author} {\bibfnamefont
  {Q.}~\bibnamefont {Si}},\ }\bibfield  {title} {\bibinfo {title} {{Shot noise
  and universal Fano factor as a characterization of strongly correlated
  metals}},\ }\href {https://doi.org/10.1103/PhysRevResearch.6.L042045}
  {\bibfield  {journal} {\bibinfo  {journal} {Phys. Rev. Res.}\ }\textbf
  {\bibinfo {volume} {6}},\ \bibinfo {pages} {L042045} (\bibinfo {year}
  {2024})}\BibitemShut {NoStop}%
\bibitem [{\citenamefont {Chen}\ \emph {et~al.}(2023)\citenamefont {Chen},
  \citenamefont {Lowder}, \citenamefont {Bakali}, \citenamefont {Andrews},
  \citenamefont {Schrenk}, \citenamefont {Waas}, \citenamefont {Svagera},
  \citenamefont {Eguchi}, \citenamefont {Prochaska}, \citenamefont {Wang} \emph
  {et~al.}}]{chen2023shot}%
  \BibitemOpen
  \bibfield  {author} {\bibinfo {author} {\bibfnamefont {L.}~\bibnamefont
  {Chen}}, \bibinfo {author} {\bibfnamefont {D.~T.}\ \bibnamefont {Lowder}},
  \bibinfo {author} {\bibfnamefont {E.}~\bibnamefont {Bakali}}, \bibinfo
  {author} {\bibfnamefont {A.~M.}\ \bibnamefont {Andrews}}, \bibinfo {author}
  {\bibfnamefont {W.}~\bibnamefont {Schrenk}}, \bibinfo {author} {\bibfnamefont
  {M.}~\bibnamefont {Waas}}, \bibinfo {author} {\bibfnamefont {R.}~\bibnamefont
  {Svagera}}, \bibinfo {author} {\bibfnamefont {G.}~\bibnamefont {Eguchi}},
  \bibinfo {author} {\bibfnamefont {L.}~\bibnamefont {Prochaska}}, \bibinfo
  {author} {\bibfnamefont {Y.}~\bibnamefont {Wang}}, \emph {et~al.},\
  }\bibfield  {title} {\bibinfo {title} {Shot noise in a strange metal},\
  }\href {https://doi.org/10.1126/science.abq6100} {\bibfield  {journal}
  {\bibinfo  {journal} {Science}\ }\textbf {\bibinfo {volume} {382}},\ \bibinfo
  {pages} {907} (\bibinfo {year} {2023})}\BibitemShut {NoStop}%
\bibitem [{\citenamefont {Nikolaenko}\ \emph {et~al.}(2023)\citenamefont
  {Nikolaenko}, \citenamefont {Sachdev},\ and\ \citenamefont
  {Patel}}]{PhysRevResearch.5.043143}%
  \BibitemOpen
  \bibfield  {author} {\bibinfo {author} {\bibfnamefont {A.}~\bibnamefont
  {Nikolaenko}}, \bibinfo {author} {\bibfnamefont {S.}~\bibnamefont
  {Sachdev}},\ and\ \bibinfo {author} {\bibfnamefont {A.~A.}\ \bibnamefont
  {Patel}},\ }\bibfield  {title} {\bibinfo {title} {Theory of shot noise in
  strange metals},\ }\href {https://doi.org/10.1103/PhysRevResearch.5.043143}
  {\bibfield  {journal} {\bibinfo  {journal} {Phys. Rev. Res.}\ }\textbf
  {\bibinfo {volume} {5}},\ \bibinfo {pages} {043143} (\bibinfo {year}
  {2023})}\BibitemShut {NoStop}%
\bibitem [{\citenamefont {Wu}\ and\ \citenamefont
  {Foster}(2024)}]{wu2023suppression}%
  \BibitemOpen
  \bibfield  {author} {\bibinfo {author} {\bibfnamefont {T.~C.}\ \bibnamefont
  {Wu}}\ and\ \bibinfo {author} {\bibfnamefont {M.~S.}\ \bibnamefont
  {Foster}},\ }\bibfield  {title} {\bibinfo {title} {{Suppression of shot noise
  in a dirty marginal Fermi liquid}},\ }\href
  {https://doi.org/10.1103/PhysRevB.110.L081102} {\bibfield  {journal}
  {\bibinfo  {journal} {Phys. Rev. B}\ }\textbf {\bibinfo {volume} {110}},\
  \bibinfo {pages} {L081102} (\bibinfo {year} {2024})}\BibitemShut {NoStop}%
\bibitem [{\citenamefont {Green}\ \emph {et~al.}(2006)\citenamefont {Green},
  \citenamefont {Moore}, \citenamefont {Sondhi},\ and\ \citenamefont
  {Vishwanath}}]{PhysRevLett.97.227003}%
  \BibitemOpen
  \bibfield  {author} {\bibinfo {author} {\bibfnamefont {A.~G.}\ \bibnamefont
  {Green}}, \bibinfo {author} {\bibfnamefont {J.~E.}\ \bibnamefont {Moore}},
  \bibinfo {author} {\bibfnamefont {S.~L.}\ \bibnamefont {Sondhi}},\ and\
  \bibinfo {author} {\bibfnamefont {A.}~\bibnamefont {Vishwanath}},\ }\bibfield
   {title} {\bibinfo {title} {{Current Noise in the Vicinity of the 2D
  Superconductor-Insulator Quantum Critical Point}},\ }\href
  {https://doi.org/10.1103/PhysRevLett.97.227003} {\bibfield  {journal}
  {\bibinfo  {journal} {Phys. Rev. Lett.}\ }\textbf {\bibinfo {volume} {97}},\
  \bibinfo {pages} {227003} (\bibinfo {year} {2006})}\BibitemShut {NoStop}%
\bibitem [{\citenamefont {Wei}\ \emph {et~al.}(2022)\citenamefont {Wei},
  \citenamefont {Rubio-Abadal}, \citenamefont {Ye}, \citenamefont {Machado},
  \citenamefont {Kemp}, \citenamefont {Srakaew}, \citenamefont {Hollerith},
  \citenamefont {Rui}, \citenamefont {Gopalakrishnan}, \citenamefont {Yao},
  \citenamefont {Bloch},\ and\ \citenamefont {Zeiher}}]{wei2022quantum}%
  \BibitemOpen
  \bibfield  {author} {\bibinfo {author} {\bibfnamefont {D.}~\bibnamefont
  {Wei}}, \bibinfo {author} {\bibfnamefont {A.}~\bibnamefont {Rubio-Abadal}},
  \bibinfo {author} {\bibfnamefont {B.}~\bibnamefont {Ye}}, \bibinfo {author}
  {\bibfnamefont {F.}~\bibnamefont {Machado}}, \bibinfo {author} {\bibfnamefont
  {J.}~\bibnamefont {Kemp}}, \bibinfo {author} {\bibfnamefont {K.}~\bibnamefont
  {Srakaew}}, \bibinfo {author} {\bibfnamefont {S.}~\bibnamefont {Hollerith}},
  \bibinfo {author} {\bibfnamefont {J.}~\bibnamefont {Rui}}, \bibinfo {author}
  {\bibfnamefont {S.}~\bibnamefont {Gopalakrishnan}}, \bibinfo {author}
  {\bibfnamefont {N.~Y.}\ \bibnamefont {Yao}}, \bibinfo {author} {\bibfnamefont
  {I.}~\bibnamefont {Bloch}},\ and\ \bibinfo {author} {\bibfnamefont
  {J.}~\bibnamefont {Zeiher}},\ }\bibfield  {title} {\bibinfo {title} {{Quantum
  gas microscopy of Kardar-Parisi-Zhang superdiffusion}},\ }\href
  {https://doi.org/10.1126/science.abk2397} {\bibfield  {journal} {\bibinfo
  {journal} {Science}\ }\textbf {\bibinfo {volume} {376}},\ \bibinfo {pages}
  {716–720} (\bibinfo {year} {2022})}\BibitemShut {NoStop}%
\bibitem [{\citenamefont {Krajnik}\ \emph
  {et~al.}(2022{\natexlab{a}})\citenamefont {Krajnik}, \citenamefont
  {Ilievski},\ and\ \citenamefont {Prosen}}]{PhysRevLett.128.090604}%
  \BibitemOpen
  \bibfield  {author} {\bibinfo {author} {\bibfnamefont {{\v{Z}}.}~\bibnamefont
  {Krajnik}}, \bibinfo {author} {\bibfnamefont {E.}~\bibnamefont {Ilievski}},\
  and\ \bibinfo {author} {\bibfnamefont {T.}~\bibnamefont {Prosen}},\
  }\bibfield  {title} {\bibinfo {title} {{Absence of Normal Fluctuations in an
  Integrable Magnet}},\ }\href {https://doi.org/10.1103/PhysRevLett.128.090604}
  {\bibfield  {journal} {\bibinfo  {journal} {Phys. Rev. Lett.}\ }\textbf
  {\bibinfo {volume} {128}},\ \bibinfo {pages} {090604} (\bibinfo {year}
  {2022}{\natexlab{a}})}\BibitemShut {NoStop}%
\bibitem [{\citenamefont {Krajnik}\ \emph
  {et~al.}(2022{\natexlab{b}})\citenamefont {Krajnik}, \citenamefont {Schmidt},
  \citenamefont {Pasquier}, \citenamefont {Ilievski},\ and\ \citenamefont
  {Prosen}}]{PhysRevLett.128.160601}%
  \BibitemOpen
  \bibfield  {author} {\bibinfo {author} {\bibfnamefont {{\v{Z}}.}~\bibnamefont
  {Krajnik}}, \bibinfo {author} {\bibfnamefont {J.}~\bibnamefont {Schmidt}},
  \bibinfo {author} {\bibfnamefont {V.}~\bibnamefont {Pasquier}}, \bibinfo
  {author} {\bibfnamefont {E.}~\bibnamefont {Ilievski}},\ and\ \bibinfo
  {author} {\bibfnamefont {T.}~\bibnamefont {Prosen}},\ }\bibfield  {title}
  {\bibinfo {title} {{Exact Anomalous Current Fluctuations in a Deterministic
  Interacting Model}},\ }\href {https://doi.org/10.1103/PhysRevLett.128.160601}
  {\bibfield  {journal} {\bibinfo  {journal} {Phys. Rev. Lett.}\ }\textbf
  {\bibinfo {volume} {128}},\ \bibinfo {pages} {160601} (\bibinfo {year}
  {2022}{\natexlab{b}})}\BibitemShut {NoStop}%
\bibitem [{\citenamefont {Gopalakrishnan}\ \emph
  {et~al.}(2024{\natexlab{a}})\citenamefont {Gopalakrishnan}, \citenamefont
  {Morningstar}, \citenamefont {Vasseur},\ and\ \citenamefont
  {Khemani}}]{PhysRevB.109.024417}%
  \BibitemOpen
  \bibfield  {author} {\bibinfo {author} {\bibfnamefont {S.}~\bibnamefont
  {Gopalakrishnan}}, \bibinfo {author} {\bibfnamefont {A.}~\bibnamefont
  {Morningstar}}, \bibinfo {author} {\bibfnamefont {R.}~\bibnamefont
  {Vasseur}},\ and\ \bibinfo {author} {\bibfnamefont {V.}~\bibnamefont
  {Khemani}},\ }\bibfield  {title} {\bibinfo {title} {Distinct universality
  classes of diffusive transport from full counting statistics},\ }\href
  {https://doi.org/10.1103/PhysRevB.109.024417} {\bibfield  {journal} {\bibinfo
   {journal} {Phys. Rev. B}\ }\textbf {\bibinfo {volume} {109}},\ \bibinfo
  {pages} {024417} (\bibinfo {year} {2024}{\natexlab{a}})}\BibitemShut
  {NoStop}%
\bibitem [{\citenamefont {McCulloch}\ \emph {et~al.}(2023)\citenamefont
  {McCulloch}, \citenamefont {De~Nardis}, \citenamefont {Gopalakrishnan},\ and\
  \citenamefont {Vasseur}}]{PhysRevLett.131.210402}%
  \BibitemOpen
  \bibfield  {author} {\bibinfo {author} {\bibfnamefont {E.}~\bibnamefont
  {McCulloch}}, \bibinfo {author} {\bibfnamefont {J.}~\bibnamefont
  {De~Nardis}}, \bibinfo {author} {\bibfnamefont {S.}~\bibnamefont
  {Gopalakrishnan}},\ and\ \bibinfo {author} {\bibfnamefont {R.}~\bibnamefont
  {Vasseur}},\ }\bibfield  {title} {\bibinfo {title} {Full counting statistics
  of charge in chaotic many-body quantum systems},\ }\href
  {https://doi.org/10.1103/PhysRevLett.131.210402} {\bibfield  {journal}
  {\bibinfo  {journal} {Phys. Rev. Lett.}\ }\textbf {\bibinfo {volume} {131}},\
  \bibinfo {pages} {210402} (\bibinfo {year} {2023})}\BibitemShut {NoStop}%
\bibitem [{\citenamefont {Rosenberg}\ \emph {et~al.}(2024)\citenamefont
  {Rosenberg}, \citenamefont {Andersen}, \citenamefont {Samajdar},
  \citenamefont {Petukhov}, \citenamefont {Hoke}, \citenamefont {Abanin},
  \citenamefont {Bengtsson}, \citenamefont {Drozdov}, \citenamefont {Erickson},
  \citenamefont {Klimov} \emph {et~al.}}]{rosenberg2023dynamics}%
  \BibitemOpen
  \bibfield  {author} {\bibinfo {author} {\bibfnamefont {E.}~\bibnamefont
  {Rosenberg}}, \bibinfo {author} {\bibfnamefont {T.}~\bibnamefont {Andersen}},
  \bibinfo {author} {\bibfnamefont {R.}~\bibnamefont {Samajdar}}, \bibinfo
  {author} {\bibfnamefont {A.}~\bibnamefont {Petukhov}}, \bibinfo {author}
  {\bibfnamefont {J.}~\bibnamefont {Hoke}}, \bibinfo {author} {\bibfnamefont
  {D.}~\bibnamefont {Abanin}}, \bibinfo {author} {\bibfnamefont
  {A.}~\bibnamefont {Bengtsson}}, \bibinfo {author} {\bibfnamefont
  {I.}~\bibnamefont {Drozdov}}, \bibinfo {author} {\bibfnamefont
  {C.}~\bibnamefont {Erickson}}, \bibinfo {author} {\bibfnamefont
  {P.}~\bibnamefont {Klimov}}, \emph {et~al.},\ }\bibfield  {title} {\bibinfo
  {title} {{Dynamics of magnetization at infinite temperature in a Heisenberg
  spin chain}},\ }\href {https://doi.org/10.1126/science.adi7877} {\bibfield
  {journal} {\bibinfo  {journal} {Science}\ }\textbf {\bibinfo {volume}
  {384}},\ \bibinfo {pages} {48} (\bibinfo {year} {2024})}\BibitemShut
  {NoStop}%
\bibitem [{\citenamefont {Wienand}\ \emph {et~al.}(2024)\citenamefont
  {Wienand}, \citenamefont {Karch}, \citenamefont {Impertro}, \citenamefont
  {Schweizer}, \citenamefont {McCulloch}, \citenamefont {Vasseur},
  \citenamefont {Gopalakrishnan}, \citenamefont {Aidelsburger},\ and\
  \citenamefont {Bloch}}]{wienand2023emergence}%
  \BibitemOpen
  \bibfield  {author} {\bibinfo {author} {\bibfnamefont {J.~F.}\ \bibnamefont
  {Wienand}}, \bibinfo {author} {\bibfnamefont {S.}~\bibnamefont {Karch}},
  \bibinfo {author} {\bibfnamefont {A.}~\bibnamefont {Impertro}}, \bibinfo
  {author} {\bibfnamefont {C.}~\bibnamefont {Schweizer}}, \bibinfo {author}
  {\bibfnamefont {E.}~\bibnamefont {McCulloch}}, \bibinfo {author}
  {\bibfnamefont {R.}~\bibnamefont {Vasseur}}, \bibinfo {author} {\bibfnamefont
  {S.}~\bibnamefont {Gopalakrishnan}}, \bibinfo {author} {\bibfnamefont
  {M.}~\bibnamefont {Aidelsburger}},\ and\ \bibinfo {author} {\bibfnamefont
  {I.}~\bibnamefont {Bloch}},\ }\bibfield  {title} {\bibinfo {title} {Emergence
  of fluctuating hydrodynamics in chaotic quantum systems},\ }\href
  {https://doi.org/10.1038/s41567-024-02611-z} {\bibfield  {journal} {\bibinfo
  {journal} {Nat. Phys.}\ }\textbf {\bibinfo {volume} {20}},\ \bibinfo {pages}
  {1732} (\bibinfo {year} {2024})}\BibitemShut {NoStop}%
\bibitem [{\citenamefont {Samajdar}\ \emph {et~al.}(2024)\citenamefont
  {Samajdar}, \citenamefont {McCulloch}, \citenamefont {Khemani}, \citenamefont
  {Vasseur},\ and\ \citenamefont {Gopalakrishnan}}]{samajdar2023quantum}%
  \BibitemOpen
  \bibfield  {author} {\bibinfo {author} {\bibfnamefont {R.}~\bibnamefont
  {Samajdar}}, \bibinfo {author} {\bibfnamefont {E.}~\bibnamefont {McCulloch}},
  \bibinfo {author} {\bibfnamefont {V.}~\bibnamefont {Khemani}}, \bibinfo
  {author} {\bibfnamefont {R.}~\bibnamefont {Vasseur}},\ and\ \bibinfo {author}
  {\bibfnamefont {S.}~\bibnamefont {Gopalakrishnan}},\ }\bibfield  {title}
  {\bibinfo {title} {{Quantum Turnstiles for Robust Measurement of Full
  Counting Statistics}},\ }\href
  {https://doi.org/10.1103/PhysRevLett.133.240403} {\bibfield  {journal}
  {\bibinfo  {journal} {Phys. Rev. Lett.}\ }\textbf {\bibinfo {volume} {133}},\
  \bibinfo {pages} {240403} (\bibinfo {year} {2024})}\BibitemShut {NoStop}%
\bibitem [{\citenamefont {Gopalakrishnan}\ \emph
  {et~al.}(2024{\natexlab{b}})\citenamefont {Gopalakrishnan}, \citenamefont
  {McCulloch},\ and\ \citenamefont {Vasseur}}]{gopalakrishnan2024non}%
  \BibitemOpen
  \bibfield  {author} {\bibinfo {author} {\bibfnamefont {S.}~\bibnamefont
  {Gopalakrishnan}}, \bibinfo {author} {\bibfnamefont {E.}~\bibnamefont
  {McCulloch}},\ and\ \bibinfo {author} {\bibfnamefont {R.}~\bibnamefont
  {Vasseur}},\ }\bibfield  {title} {\bibinfo {title} {{Non-Gaussian diffusive
  fluctuations in Dirac fluids}},\ }\href
  {https://doi.org/10.1073/pnas.2403327121} {\bibfield  {journal} {\bibinfo
  {journal} {Proc. Natl. Acad. Sci. U.S.A.}\ }\textbf {\bibinfo {volume}
  {121}},\ \bibinfo {pages} {e2403327121} (\bibinfo {year}
  {2024}{\natexlab{b}})}\BibitemShut {NoStop}%
\bibitem [{\citenamefont {Blanter}\ and\ \citenamefont
  {B{\"u}ttiker}(2000)}]{blanter2000shot}%
  \BibitemOpen
  \bibfield  {author} {\bibinfo {author} {\bibfnamefont {Y.~M.}\ \bibnamefont
  {Blanter}}\ and\ \bibinfo {author} {\bibfnamefont {M.}~\bibnamefont
  {B{\"u}ttiker}},\ }\bibfield  {title} {\bibinfo {title} {Shot noise in
  mesoscopic conductors},\ }\href
  {https://doi.org/10.1016/S0370-1573(99)00123-4} {\bibfield  {journal}
  {\bibinfo  {journal} {Phys. Rep.}\ }\textbf {\bibinfo {volume} {336}},\
  \bibinfo {pages} {1} (\bibinfo {year} {2000})}\BibitemShut {NoStop}%
\bibitem [{\citenamefont {Tikhonov}\ \emph {et~al.}(2020)\citenamefont
  {Tikhonov}, \citenamefont {Denisov}, \citenamefont {Piatrusha}, \citenamefont
  {Khrapach}, \citenamefont {Pekola}, \citenamefont {Karimi}, \citenamefont
  {Jabdaraghi},\ and\ \citenamefont {Khrapai}}]{tikhonov2020spatial}%
  \BibitemOpen
  \bibfield  {author} {\bibinfo {author} {\bibfnamefont {E.~S.}\ \bibnamefont
  {Tikhonov}}, \bibinfo {author} {\bibfnamefont {A.~O.}\ \bibnamefont
  {Denisov}}, \bibinfo {author} {\bibfnamefont {S.~U.}\ \bibnamefont
  {Piatrusha}}, \bibinfo {author} {\bibfnamefont {I.~N.}\ \bibnamefont
  {Khrapach}}, \bibinfo {author} {\bibfnamefont {J.~P.}\ \bibnamefont
  {Pekola}}, \bibinfo {author} {\bibfnamefont {B.}~\bibnamefont {Karimi}},
  \bibinfo {author} {\bibfnamefont {R.~N.}\ \bibnamefont {Jabdaraghi}},\ and\
  \bibinfo {author} {\bibfnamefont {V.~S.}\ \bibnamefont {Khrapai}},\
  }\bibfield  {title} {\bibinfo {title} {Spatial and energy resolution of
  electronic states by shot noise},\ }\href
  {https://doi.org/10.1103/PhysRevB.102.085417} {\bibfield  {journal} {\bibinfo
   {journal} {Phys. Rev. B}\ }\textbf {\bibinfo {volume} {102}},\ \bibinfo
  {pages} {085417} (\bibinfo {year} {2020})}\BibitemShut {NoStop}%
\bibitem [{\citenamefont {Pelliccione}\ \emph {et~al.}(2016)\citenamefont
  {Pelliccione}, \citenamefont {Jenkins}, \citenamefont {Ovartchaiyapong},
  \citenamefont {Reetz}, \citenamefont {Emmanouilidou}, \citenamefont {Ni},\
  and\ \citenamefont {Bleszynski~Jayich}}]{Pelliccione2016a}%
  \BibitemOpen
  \bibfield  {author} {\bibinfo {author} {\bibfnamefont {M.}~\bibnamefont
  {Pelliccione}}, \bibinfo {author} {\bibfnamefont {A.}~\bibnamefont
  {Jenkins}}, \bibinfo {author} {\bibfnamefont {P.}~\bibnamefont
  {Ovartchaiyapong}}, \bibinfo {author} {\bibfnamefont {C.}~\bibnamefont
  {Reetz}}, \bibinfo {author} {\bibfnamefont {E.}~\bibnamefont
  {Emmanouilidou}}, \bibinfo {author} {\bibfnamefont {N.}~\bibnamefont {Ni}},\
  and\ \bibinfo {author} {\bibfnamefont {A.~C.}\ \bibnamefont
  {Bleszynski~Jayich}},\ }\bibfield  {title} {\bibinfo {title} {Scanned probe
  imaging of nanoscale magnetism at cryogenic temperatures with a single-spin
  quantum sensor},\ }\href {https://doi.org/10.1038/nnano.2016.68} {\bibfield
  {journal} {\bibinfo  {journal} {Nat. Nanotech.}\ }\textbf {\bibinfo {volume}
  {11}},\ \bibinfo {pages} {700} (\bibinfo {year} {2016})}\BibitemShut
  {NoStop}%
\bibitem [{\citenamefont {Ariyaratne}\ \emph {et~al.}(2018)\citenamefont
  {Ariyaratne}, \citenamefont {Bluvstein}, \citenamefont {Myers},\ and\
  \citenamefont {Jayich}}]{Ariyaratne2018a}%
  \BibitemOpen
  \bibfield  {author} {\bibinfo {author} {\bibfnamefont {A.}~\bibnamefont
  {Ariyaratne}}, \bibinfo {author} {\bibfnamefont {D.}~\bibnamefont
  {Bluvstein}}, \bibinfo {author} {\bibfnamefont {B.~A.}\ \bibnamefont
  {Myers}},\ and\ \bibinfo {author} {\bibfnamefont {A.~C.~B.}\ \bibnamefont
  {Jayich}},\ }\bibfield  {title} {\bibinfo {title} {Nanoscale electrical
  conductivity imaging using a nitrogen-vacancy center in diamond},\ }\href
  {https://doi.org/10.1038/s41467-018-04798-1} {\bibfield  {journal} {\bibinfo
  {journal} {Nat. Commun.}\ }\textbf {\bibinfo {volume} {9}},\ \bibinfo {pages}
  {2406} (\bibinfo {year} {2018})}\BibitemShut {NoStop}%
\bibitem [{\citenamefont {Rodriguez-Nieva}\ \emph {et~al.}(2018)\citenamefont
  {Rodriguez-Nieva}, \citenamefont {Agarwal}, \citenamefont {Giamarchi},
  \citenamefont {Halperin}, \citenamefont {Lukin},\ and\ \citenamefont
  {Demler}}]{PhysRevB.98.195433}%
  \BibitemOpen
  \bibfield  {author} {\bibinfo {author} {\bibfnamefont {J.~F.}\ \bibnamefont
  {Rodriguez-Nieva}}, \bibinfo {author} {\bibfnamefont {K.}~\bibnamefont
  {Agarwal}}, \bibinfo {author} {\bibfnamefont {T.}~\bibnamefont {Giamarchi}},
  \bibinfo {author} {\bibfnamefont {B.~I.}\ \bibnamefont {Halperin}}, \bibinfo
  {author} {\bibfnamefont {M.~D.}\ \bibnamefont {Lukin}},\ and\ \bibinfo
  {author} {\bibfnamefont {E.}~\bibnamefont {Demler}},\ }\bibfield  {title}
  {\bibinfo {title} {Probing one-dimensional systems via noise magnetometry
  with single spin qubits},\ }\href
  {https://doi.org/10.1103/PhysRevB.98.195433} {\bibfield  {journal} {\bibinfo
  {journal} {Phys. Rev. B}\ }\textbf {\bibinfo {volume} {98}},\ \bibinfo
  {pages} {195433} (\bibinfo {year} {2018})}\BibitemShut {NoStop}%
\bibitem [{\citenamefont {Andersen}\ \emph {et~al.}(2019)\citenamefont
  {Andersen}, \citenamefont {Dwyer}, \citenamefont {Sanchez-Yamagishi},
  \citenamefont {Rodriguez-Nieva}, \citenamefont {Agarwal}, \citenamefont
  {Watanabe}, \citenamefont {Taniguchi}, \citenamefont {Demler}, \citenamefont
  {Kim}, \citenamefont {Park},\ and\ \citenamefont {Lukin}}]{Andersen2019a}%
  \BibitemOpen
  \bibfield  {author} {\bibinfo {author} {\bibfnamefont {T.~I.}\ \bibnamefont
  {Andersen}}, \bibinfo {author} {\bibfnamefont {B.~L.}\ \bibnamefont {Dwyer}},
  \bibinfo {author} {\bibfnamefont {J.~D.}\ \bibnamefont {Sanchez-Yamagishi}},
  \bibinfo {author} {\bibfnamefont {J.~F.}\ \bibnamefont {Rodriguez-Nieva}},
  \bibinfo {author} {\bibfnamefont {K.}~\bibnamefont {Agarwal}}, \bibinfo
  {author} {\bibfnamefont {K.}~\bibnamefont {Watanabe}}, \bibinfo {author}
  {\bibfnamefont {T.}~\bibnamefont {Taniguchi}}, \bibinfo {author}
  {\bibfnamefont {E.~A.}\ \bibnamefont {Demler}}, \bibinfo {author}
  {\bibfnamefont {P.}~\bibnamefont {Kim}}, \bibinfo {author} {\bibfnamefont
  {H.}~\bibnamefont {Park}},\ and\ \bibinfo {author} {\bibfnamefont {M.~D.}\
  \bibnamefont {Lukin}},\ }\bibfield  {title} {\bibinfo {title}
  {Electron-phonon instability in graphene revealed by global and local noise
  probes},\ }\href {https://doi.org/10.1126/science.aaw2104} {\bibfield
  {journal} {\bibinfo  {journal} {Science}\ }\textbf {\bibinfo {volume}
  {364}},\ \bibinfo {pages} {154} (\bibinfo {year} {2019})}\BibitemShut
  {NoStop}%
\bibitem [{\citenamefont {Lee-Wong}\ \emph {et~al.}(2020)\citenamefont
  {Lee-Wong}, \citenamefont {Xue}, \citenamefont {Ye}, \citenamefont {Kreisel},
  \citenamefont {Van Der~Sar}, \citenamefont {Yacoby},\ and\ \citenamefont
  {Du}}]{Lee-Wong2020}%
  \BibitemOpen
  \bibfield  {author} {\bibinfo {author} {\bibfnamefont {E.}~\bibnamefont
  {Lee-Wong}}, \bibinfo {author} {\bibfnamefont {R.}~\bibnamefont {Xue}},
  \bibinfo {author} {\bibfnamefont {F.}~\bibnamefont {Ye}}, \bibinfo {author}
  {\bibfnamefont {A.}~\bibnamefont {Kreisel}}, \bibinfo {author} {\bibfnamefont
  {T.}~\bibnamefont {Van Der~Sar}}, \bibinfo {author} {\bibfnamefont
  {A.}~\bibnamefont {Yacoby}},\ and\ \bibinfo {author} {\bibfnamefont {C.~R.}\
  \bibnamefont {Du}},\ }\bibfield  {title} {\bibinfo {title} {Nanoscale
  {Detection} of {Magnon} {Excitations} with {Variable} {Wavevectors} {Through}
  a {Quantum} {Spin} {Sensor}},\ }\href
  {https://doi.org/10.1021/acs.nanolett.0c00085} {\bibfield  {journal}
  {\bibinfo  {journal} {Nano Lett.}\ }\textbf {\bibinfo {volume} {20}},\
  \bibinfo {pages} {3284} (\bibinfo {year} {2020})}\BibitemShut {NoStop}%
\bibitem [{\citenamefont {Ku}\ \emph {et~al.}(2020)\citenamefont {Ku},
  \citenamefont {Zhou}, \citenamefont {Li}, \citenamefont {Shin}, \citenamefont
  {Shi}, \citenamefont {Burch}, \citenamefont {Anderson}, \citenamefont
  {Pierce}, \citenamefont {Xie}, \citenamefont {Hamo}, \citenamefont {Vool},
  \citenamefont {Zhang}, \citenamefont {Casola}, \citenamefont {Taniguchi},
  \citenamefont {Watanabe}, \citenamefont {Fogler}, \citenamefont {Kim},
  \citenamefont {Yacoby},\ and\ \citenamefont
  {Walsworth}}]{10.1038/s41586-020-2507-2}%
  \BibitemOpen
  \bibfield  {author} {\bibinfo {author} {\bibfnamefont {M.~J.}\ \bibnamefont
  {Ku}}, \bibinfo {author} {\bibfnamefont {T.~X.}\ \bibnamefont {Zhou}},
  \bibinfo {author} {\bibfnamefont {Q.}~\bibnamefont {Li}}, \bibinfo {author}
  {\bibfnamefont {Y.~J.}\ \bibnamefont {Shin}}, \bibinfo {author}
  {\bibfnamefont {J.~K.}\ \bibnamefont {Shi}}, \bibinfo {author} {\bibfnamefont
  {C.}~\bibnamefont {Burch}}, \bibinfo {author} {\bibfnamefont {L.~E.}\
  \bibnamefont {Anderson}}, \bibinfo {author} {\bibfnamefont {A.~T.}\
  \bibnamefont {Pierce}}, \bibinfo {author} {\bibfnamefont {Y.}~\bibnamefont
  {Xie}}, \bibinfo {author} {\bibfnamefont {A.}~\bibnamefont {Hamo}}, \bibinfo
  {author} {\bibfnamefont {U.}~\bibnamefont {Vool}}, \bibinfo {author}
  {\bibfnamefont {H.}~\bibnamefont {Zhang}}, \bibinfo {author} {\bibfnamefont
  {F.}~\bibnamefont {Casola}}, \bibinfo {author} {\bibfnamefont
  {T.}~\bibnamefont {Taniguchi}}, \bibinfo {author} {\bibfnamefont
  {K.}~\bibnamefont {Watanabe}}, \bibinfo {author} {\bibfnamefont {M.~M.}\
  \bibnamefont {Fogler}}, \bibinfo {author} {\bibfnamefont {P.}~\bibnamefont
  {Kim}}, \bibinfo {author} {\bibfnamefont {A.}~\bibnamefont {Yacoby}},\ and\
  \bibinfo {author} {\bibfnamefont {R.~L.}\ \bibnamefont {Walsworth}},\
  }\bibfield  {title} {\bibinfo {title} {{Imaging viscous flow of the Dirac
  fluid in graphene}},\ }\href {https://doi.org/10.1038/s41586-020-2507-2}
  {\bibfield  {journal} {\bibinfo  {journal} {Nature}\ }\textbf {\bibinfo
  {volume} {583}},\ \bibinfo {pages} {537} (\bibinfo {year} {2020})},\ \Eprint
  {https://arxiv.org/abs/1905.10791} {1905.10791} \BibitemShut {NoStop}%
\bibitem [{\citenamefont {Zhou}\ \emph {et~al.}(2021)\citenamefont {Zhou},
  \citenamefont {Carmiggelt}, \citenamefont {Gächter}, \citenamefont
  {Esterlis}, \citenamefont {Sels}, \citenamefont {Stöhr}, \citenamefont {Du},
  \citenamefont {Fernandez}, \citenamefont {Rodriguez-Nieva}, \citenamefont
  {Büttner}, \citenamefont {Demler},\ and\ \citenamefont
  {Yacoby}}]{Zhou2021a}%
  \BibitemOpen
  \bibfield  {author} {\bibinfo {author} {\bibfnamefont {T.~X.}\ \bibnamefont
  {Zhou}}, \bibinfo {author} {\bibfnamefont {J.~J.}\ \bibnamefont
  {Carmiggelt}}, \bibinfo {author} {\bibfnamefont {L.~M.}\ \bibnamefont
  {Gächter}}, \bibinfo {author} {\bibfnamefont {I.}~\bibnamefont {Esterlis}},
  \bibinfo {author} {\bibfnamefont {D.}~\bibnamefont {Sels}}, \bibinfo {author}
  {\bibfnamefont {R.~J.}\ \bibnamefont {Stöhr}}, \bibinfo {author}
  {\bibfnamefont {C.}~\bibnamefont {Du}}, \bibinfo {author} {\bibfnamefont
  {D.}~\bibnamefont {Fernandez}}, \bibinfo {author} {\bibfnamefont {J.~F.}\
  \bibnamefont {Rodriguez-Nieva}}, \bibinfo {author} {\bibfnamefont
  {F.}~\bibnamefont {Büttner}}, \bibinfo {author} {\bibfnamefont
  {E.}~\bibnamefont {Demler}},\ and\ \bibinfo {author} {\bibfnamefont
  {A.}~\bibnamefont {Yacoby}},\ }\bibfield  {title} {\bibinfo {title} {A magnon
  scattering platform},\ }\href {https://doi.org/10.1073/pnas.2019473118}
  {\bibfield  {journal} {\bibinfo  {journal} {Proc. Natl. Acad. Sci. U.S.A.}\
  }\textbf {\bibinfo {volume} {118}},\ \bibinfo {pages} {e2019473118} (\bibinfo
  {year} {2021})}\BibitemShut {NoStop}%
\bibitem [{\citenamefont {Vool}\ \emph {et~al.}(2021)\citenamefont {Vool},
  \citenamefont {Hamo}, \citenamefont {Varnavides}, \citenamefont {Wang},
  \citenamefont {Zhou}, \citenamefont {Kumar}, \citenamefont {Dovzhenko},
  \citenamefont {Qiu}, \citenamefont {Garcia}, \citenamefont {Pierce},
  \citenamefont {Gooth}, \citenamefont {Anikeeva}, \citenamefont {Felser},
  \citenamefont {Narang},\ and\ \citenamefont
  {Yacoby}}]{10.1038/s41567-021-01341-w}%
  \BibitemOpen
  \bibfield  {author} {\bibinfo {author} {\bibfnamefont {U.}~\bibnamefont
  {Vool}}, \bibinfo {author} {\bibfnamefont {A.}~\bibnamefont {Hamo}}, \bibinfo
  {author} {\bibfnamefont {G.}~\bibnamefont {Varnavides}}, \bibinfo {author}
  {\bibfnamefont {Y.}~\bibnamefont {Wang}}, \bibinfo {author} {\bibfnamefont
  {T.~X.}\ \bibnamefont {Zhou}}, \bibinfo {author} {\bibfnamefont
  {N.}~\bibnamefont {Kumar}}, \bibinfo {author} {\bibfnamefont
  {Y.}~\bibnamefont {Dovzhenko}}, \bibinfo {author} {\bibfnamefont
  {Z.}~\bibnamefont {Qiu}}, \bibinfo {author} {\bibfnamefont {C.~A.}\
  \bibnamefont {Garcia}}, \bibinfo {author} {\bibfnamefont {A.~T.}\
  \bibnamefont {Pierce}}, \bibinfo {author} {\bibfnamefont {J.}~\bibnamefont
  {Gooth}}, \bibinfo {author} {\bibfnamefont {P.}~\bibnamefont {Anikeeva}},
  \bibinfo {author} {\bibfnamefont {C.}~\bibnamefont {Felser}}, \bibinfo
  {author} {\bibfnamefont {P.}~\bibnamefont {Narang}},\ and\ \bibinfo {author}
  {\bibfnamefont {A.}~\bibnamefont {Yacoby}},\ }\bibfield  {title} {\bibinfo
  {title} {{Imaging phonon-mediated hydrodynamic flow in WTe$_2$}},\ }\href
  {https://doi.org/10.1038/s41567-021-01341-w} {\bibfield  {journal} {\bibinfo
  {journal} {Nat. Phys.}\ }\textbf {\bibinfo {volume} {17}},\ \bibinfo {pages}
  {1216} (\bibinfo {year} {2021})}\BibitemShut {NoStop}%
\bibitem [{\citenamefont {Jenkins}\ \emph {et~al.}(2022)\citenamefont
  {Jenkins}, \citenamefont {Baumann}, \citenamefont {Zhou}, \citenamefont
  {Meynell}, \citenamefont {Daipeng}, \citenamefont {Watanabe}, \citenamefont
  {Taniguchi}, \citenamefont {Lucas}, \citenamefont {Young},\ and\
  \citenamefont {Jayich}}]{10.1103/physrevlett.129.087701}%
  \BibitemOpen
  \bibfield  {author} {\bibinfo {author} {\bibfnamefont {A.}~\bibnamefont
  {Jenkins}}, \bibinfo {author} {\bibfnamefont {S.}~\bibnamefont {Baumann}},
  \bibinfo {author} {\bibfnamefont {H.}~\bibnamefont {Zhou}}, \bibinfo {author}
  {\bibfnamefont {S.~A.}\ \bibnamefont {Meynell}}, \bibinfo {author}
  {\bibfnamefont {Y.}~\bibnamefont {Daipeng}}, \bibinfo {author} {\bibfnamefont
  {K.}~\bibnamefont {Watanabe}}, \bibinfo {author} {\bibfnamefont
  {T.}~\bibnamefont {Taniguchi}}, \bibinfo {author} {\bibfnamefont
  {A.}~\bibnamefont {Lucas}}, \bibinfo {author} {\bibfnamefont {A.~F.}\
  \bibnamefont {Young}},\ and\ \bibinfo {author} {\bibfnamefont {A.~C.~B.}\
  \bibnamefont {Jayich}},\ }\bibfield  {title} {\bibinfo {title} {{Imaging the
  Breakdown of Ohmic Transport in Graphene}},\ }\href
  {https://doi.org/10.1103/physrevlett.129.087701} {\bibfield  {journal}
  {\bibinfo  {journal} {Phys. Rev. Lett.}\ }\textbf {\bibinfo {volume} {129}},\
  \bibinfo {pages} {087701} (\bibinfo {year} {2022})}\BibitemShut {NoStop}%
\bibitem [{\citenamefont {Rovny}\ \emph {et~al.}(2022)\citenamefont {Rovny},
  \citenamefont {Yuan}, \citenamefont {Fitzpatrick}, \citenamefont {Abdalla},
  \citenamefont {Futamura}, \citenamefont {Fox}, \citenamefont {Cambria},
  \citenamefont {Kolkowitz},\ and\ \citenamefont {De~Leon}}]{Rovny2022a}%
  \BibitemOpen
  \bibfield  {author} {\bibinfo {author} {\bibfnamefont {J.}~\bibnamefont
  {Rovny}}, \bibinfo {author} {\bibfnamefont {Z.}~\bibnamefont {Yuan}},
  \bibinfo {author} {\bibfnamefont {M.}~\bibnamefont {Fitzpatrick}}, \bibinfo
  {author} {\bibfnamefont {A.~I.}\ \bibnamefont {Abdalla}}, \bibinfo {author}
  {\bibfnamefont {L.}~\bibnamefont {Futamura}}, \bibinfo {author}
  {\bibfnamefont {C.}~\bibnamefont {Fox}}, \bibinfo {author} {\bibfnamefont
  {M.~C.}\ \bibnamefont {Cambria}}, \bibinfo {author} {\bibfnamefont
  {S.}~\bibnamefont {Kolkowitz}},\ and\ \bibinfo {author} {\bibfnamefont
  {N.~P.}\ \bibnamefont {De~Leon}},\ }\bibfield  {title} {\bibinfo {title}
  {Nanoscale covariance magnetometry with diamond quantum sensors},\ }\href
  {https://doi.org/10.1126/science.ade9858} {\bibfield  {journal} {\bibinfo
  {journal} {Science}\ }\textbf {\bibinfo {volume} {378}},\ \bibinfo {pages}
  {1301} (\bibinfo {year} {2022})}\BibitemShut {NoStop}%
\bibitem [{\citenamefont {Rovny}\ \emph {et~al.}(2024)\citenamefont {Rovny},
  \citenamefont {Gopalakrishnan}, \citenamefont {Jayich}, \citenamefont
  {Maletinsky}, \citenamefont {Demler},\ and\ \citenamefont
  {de~Leon}}]{rovny2024new}%
  \BibitemOpen
  \bibfield  {author} {\bibinfo {author} {\bibfnamefont {J.}~\bibnamefont
  {Rovny}}, \bibinfo {author} {\bibfnamefont {S.}~\bibnamefont
  {Gopalakrishnan}}, \bibinfo {author} {\bibfnamefont {A.~C.~B.}\ \bibnamefont
  {Jayich}}, \bibinfo {author} {\bibfnamefont {P.}~\bibnamefont {Maletinsky}},
  \bibinfo {author} {\bibfnamefont {E.}~\bibnamefont {Demler}},\ and\ \bibinfo
  {author} {\bibfnamefont {N.~P.}\ \bibnamefont {de~Leon}},\ }\bibfield
  {title} {\bibinfo {title} {Nanoscale diamond quantum sensors for many-body
  physics},\ }\href {https://doi.org/10.1038/s42254-024-00775-4} {\bibfield
  {journal} {\bibinfo  {journal} {Nat. Rev. Phys.}\ }\textbf {\bibinfo {volume}
  {6}},\ \bibinfo {pages} {753} (\bibinfo {year} {2024})}\BibitemShut {NoStop}%
\bibitem [{\citenamefont {Cheng}\ \emph {et~al.}(2025)\citenamefont {Cheng},
  \citenamefont {Kazi}, \citenamefont {Rovny}, \citenamefont {Zhang},
  \citenamefont {Nassar}, \citenamefont {Thompson},\ and\ \citenamefont
  {de~Leon}}]{cheng2025massively}%
  \BibitemOpen
  \bibfield  {author} {\bibinfo {author} {\bibfnamefont {K.-H.}\ \bibnamefont
  {Cheng}}, \bibinfo {author} {\bibfnamefont {Z.}~\bibnamefont {Kazi}},
  \bibinfo {author} {\bibfnamefont {J.}~\bibnamefont {Rovny}}, \bibinfo
  {author} {\bibfnamefont {B.}~\bibnamefont {Zhang}}, \bibinfo {author}
  {\bibfnamefont {L.~S.}\ \bibnamefont {Nassar}}, \bibinfo {author}
  {\bibfnamefont {J.~D.}\ \bibnamefont {Thompson}},\ and\ \bibinfo {author}
  {\bibfnamefont {N.~P.}\ \bibnamefont {de~Leon}},\ }\bibfield  {title}
  {\bibinfo {title} {{Massively Multiplexed Nanoscale Magnetometry with Diamond
  Quantum Sensors}},\ }\href {https://doi.org/10.1103/t8fz-3tzs} {\bibfield
  {journal} {\bibinfo  {journal} {Phys. Rev. X}\ }\textbf {\bibinfo {volume}
  {15}},\ \bibinfo {pages} {031014} (\bibinfo {year} {2025})}\BibitemShut
  {NoStop}%
\bibitem [{\citenamefont {Cambria}\ \emph {et~al.}(2025)\citenamefont
  {Cambria}, \citenamefont {Chand}, \citenamefont {Reiter},\ and\ \citenamefont
  {Kolkowitz}}]{cambria2025scalable}%
  \BibitemOpen
  \bibfield  {author} {\bibinfo {author} {\bibfnamefont {M.}~\bibnamefont
  {Cambria}}, \bibinfo {author} {\bibfnamefont {S.}~\bibnamefont {Chand}},
  \bibinfo {author} {\bibfnamefont {C.~M.}\ \bibnamefont {Reiter}},\ and\
  \bibinfo {author} {\bibfnamefont {S.}~\bibnamefont {Kolkowitz}},\ }\bibfield
  {title} {\bibinfo {title} {{Scalable Parallel Measurement of Individual
  Nitrogen-Vacancy Centers}},\ }\href {https://doi.org/10.1103/jdzq-jbfz}
  {\bibfield  {journal} {\bibinfo  {journal} {Phys. Rev. X}\ }\textbf {\bibinfo
  {volume} {15}},\ \bibinfo {pages} {031015} (\bibinfo {year}
  {2025})}\BibitemShut {NoStop}%
\bibitem [{\citenamefont {Gullans}\ and\ \citenamefont
  {Huse}(2019)}]{PhysRevLett.123.110601}%
  \BibitemOpen
  \bibfield  {author} {\bibinfo {author} {\bibfnamefont {M.~J.}\ \bibnamefont
  {Gullans}}\ and\ \bibinfo {author} {\bibfnamefont {D.~A.}\ \bibnamefont
  {Huse}},\ }\bibfield  {title} {\bibinfo {title} {{Localization as an
  Entanglement Phase Transition in Boundary-Driven Anderson Models}},\ }\href
  {https://doi.org/10.1103/PhysRevLett.123.110601} {\bibfield  {journal}
  {\bibinfo  {journal} {Phys. Rev. Lett.}\ }\textbf {\bibinfo {volume} {123}},\
  \bibinfo {pages} {110601} (\bibinfo {year} {2019})}\BibitemShut {NoStop}%
\bibitem [{\citenamefont {Hruza}\ and\ \citenamefont
  {Bernard}(2023)}]{PhysRevX.13.011045}%
  \BibitemOpen
  \bibfield  {author} {\bibinfo {author} {\bibfnamefont {L.}~\bibnamefont
  {Hruza}}\ and\ \bibinfo {author} {\bibfnamefont {D.}~\bibnamefont
  {Bernard}},\ }\bibfield  {title} {\bibinfo {title} {{Coherent Fluctuations in
  Noisy Mesoscopic Systems, the Open Quantum SSEP, and Free Probability}},\
  }\href {https://doi.org/10.1103/PhysRevX.13.011045} {\bibfield  {journal}
  {\bibinfo  {journal} {Phys. Rev. X}\ }\textbf {\bibinfo {volume} {13}},\
  \bibinfo {pages} {011045} (\bibinfo {year} {2023})}\BibitemShut {NoStop}%
\bibitem [{\citenamefont {Kogan}\ and\ \citenamefont
  {Shul’man}(1969)}]{kogan1969theory}%
  \BibitemOpen
  \bibfield  {author} {\bibinfo {author} {\bibfnamefont {S.~M.}\ \bibnamefont
  {Kogan}}\ and\ \bibinfo {author} {\bibfnamefont {A.~Y.}\ \bibnamefont
  {Shul’man}},\ }\bibfield  {title} {\bibinfo {title} {Theory of fluctuations
  in a nonequilibrium electron gas},\ }\href
  {http://www.jetp.ras.ru/cgi-bin/dn/e_029_03_0467.pdf} {\bibfield  {journal}
  {\bibinfo  {journal} {Sov. Phys. JETP}\ }\textbf {\bibinfo {volume} {29}},\
  \bibinfo {pages} {467} (\bibinfo {year} {1969})}\BibitemShut {NoStop}%
\bibitem [{\citenamefont {Bixon}\ and\ \citenamefont
  {Zwanzig}(1969)}]{PhysRev.187.267}%
  \BibitemOpen
  \bibfield  {author} {\bibinfo {author} {\bibfnamefont {M.}~\bibnamefont
  {Bixon}}\ and\ \bibinfo {author} {\bibfnamefont {R.}~\bibnamefont
  {Zwanzig}},\ }\bibfield  {title} {\bibinfo {title} {{Boltzmann-Langevin
  Equation and Hydrodynamic Fluctuations}},\ }\href
  {https://doi.org/10.1103/PhysRev.187.267} {\bibfield  {journal} {\bibinfo
  {journal} {Phys. Rev.}\ }\textbf {\bibinfo {volume} {187}},\ \bibinfo {pages}
  {267} (\bibinfo {year} {1969})}\BibitemShut {NoStop}%
\bibitem [{\citenamefont {Beenakker}\ and\ \citenamefont
  {B\"uttiker}(1992)}]{PhysRevB.46.1889}%
  \BibitemOpen
  \bibfield  {author} {\bibinfo {author} {\bibfnamefont {C.~W.~J.}\
  \bibnamefont {Beenakker}}\ and\ \bibinfo {author} {\bibfnamefont
  {M.}~\bibnamefont {B\"uttiker}},\ }\bibfield  {title} {\bibinfo {title}
  {Suppression of shot noise in metallic diffusive conductors},\ }\href
  {https://doi.org/10.1103/PhysRevB.46.1889} {\bibfield  {journal} {\bibinfo
  {journal} {Phys. Rev. B}\ }\textbf {\bibinfo {volume} {46}},\ \bibinfo
  {pages} {1889} (\bibinfo {year} {1992})}\BibitemShut {NoStop}%
\bibitem [{\citenamefont {Gantsevich}\ \emph {et~al.}(1979)\citenamefont
  {Gantsevich}, \citenamefont {Gurevich},\ and\ \citenamefont
  {Katilius}}]{gantsevich1979theory}%
  \BibitemOpen
  \bibfield  {author} {\bibinfo {author} {\bibfnamefont {S.}~\bibnamefont
  {Gantsevich}}, \bibinfo {author} {\bibfnamefont {V.~L.}\ \bibnamefont
  {Gurevich}},\ and\ \bibinfo {author} {\bibfnamefont {R.}~\bibnamefont
  {Katilius}},\ }\bibfield  {title} {\bibinfo {title} {Theory of fluctuations
  in nonequilibrium electron gas},\ }\href {https://doi.org/10.1007/BF02724353}
  {\bibfield  {journal} {\bibinfo  {journal} {Riv. Nuovo Cim.}\ }\textbf
  {\bibinfo {volume} {2}},\ \bibinfo {pages} {1} (\bibinfo {year}
  {1979})}\BibitemShut {NoStop}%
\bibitem [{\citenamefont {Pitaevskii}\ and\ \citenamefont
  {Lifshitz}(2012)}]{pitaevskii2012physical}%
  \BibitemOpen
  \bibfield  {author} {\bibinfo {author} {\bibfnamefont {L.~P.}\ \bibnamefont
  {Pitaevskii}}\ and\ \bibinfo {author} {\bibfnamefont {E.~M.}\ \bibnamefont
  {Lifshitz}},\ }\href@noop {} {\emph {\bibinfo {title} {Physical Kinetics}}},\
  \bibinfo {series} {Course of Theoretical Physics}, Vol.~\bibinfo {volume}
  {10}\ (\bibinfo  {publisher} {Butterworth-Heinemann},\ \bibinfo {year}
  {2012})\BibitemShut {NoStop}%
\bibitem [{\citenamefont {Forster}(2018)}]{forster2018hydrodynamic}%
  \BibitemOpen
  \bibfield  {author} {\bibinfo {author} {\bibfnamefont {D.}~\bibnamefont
  {Forster}},\ }\href {https://doi.org/10.1201/9780429493683} {\emph {\bibinfo
  {title} {{Hydrodynamic Fluctuations, Broken Symmetry, And Correlation
  Functions}}}}\ (\bibinfo  {publisher} {CRC Press},\ \bibinfo {address} {Boca
  Raton},\ \bibinfo {year} {2018})\BibitemShut {NoStop}%
\bibitem [{\citenamefont {Barry}\ \emph {et~al.}(2020)\citenamefont {Barry},
  \citenamefont {Schloss}, \citenamefont {Bauch}, \citenamefont {Turner},
  \citenamefont {Hart}, \citenamefont {Pham},\ and\ \citenamefont
  {Walsworth}}]{barry2020}%
  \BibitemOpen
  \bibfield  {author} {\bibinfo {author} {\bibfnamefont {J.~F.}\ \bibnamefont
  {Barry}}, \bibinfo {author} {\bibfnamefont {J.~M.}\ \bibnamefont {Schloss}},
  \bibinfo {author} {\bibfnamefont {E.}~\bibnamefont {Bauch}}, \bibinfo
  {author} {\bibfnamefont {M.~J.}\ \bibnamefont {Turner}}, \bibinfo {author}
  {\bibfnamefont {C.~A.}\ \bibnamefont {Hart}}, \bibinfo {author}
  {\bibfnamefont {L.~M.}\ \bibnamefont {Pham}},\ and\ \bibinfo {author}
  {\bibfnamefont {R.~L.}\ \bibnamefont {Walsworth}},\ }\bibfield  {title}
  {\bibinfo {title} {{Sensitivity optimization for NV-diamond magnetometry}},\
  }\href {https://doi.org/10.1103/revmodphys.92.015004} {\bibfield  {journal}
  {\bibinfo  {journal} {Rev. Mod. Phys.}\ }\textbf {\bibinfo {volume} {92}},\
  \bibinfo {pages} {015004} (\bibinfo {year} {2020})},\ \Eprint
  {https://arxiv.org/abs/1903.08176} {1903.08176} \BibitemShut {NoStop}%
\bibitem [{\citenamefont {Machado}\ \emph {et~al.}(2023)\citenamefont
  {Machado}, \citenamefont {Demler}, \citenamefont {Yao},\ and\ \citenamefont
  {Chatterjee}}]{machado2022}%
  \BibitemOpen
  \bibfield  {author} {\bibinfo {author} {\bibfnamefont {F.}~\bibnamefont
  {Machado}}, \bibinfo {author} {\bibfnamefont {E.~A.}\ \bibnamefont {Demler}},
  \bibinfo {author} {\bibfnamefont {N.~Y.}\ \bibnamefont {Yao}},\ and\ \bibinfo
  {author} {\bibfnamefont {S.}~\bibnamefont {Chatterjee}},\ }\bibfield  {title}
  {\bibinfo {title} {{Quantum Noise Spectroscopy of Dynamical Critical
  Phenomena}},\ }\href {https://doi.org/10.1103/PhysRevLett.131.070801}
  {\bibfield  {journal} {\bibinfo  {journal} {Phys. Rev. Lett.}\ }\textbf
  {\bibinfo {volume} {131}},\ \bibinfo {pages} {070801} (\bibinfo {year}
  {2023})}\BibitemShut {NoStop}%
\bibitem [{\citenamefont {Nagaev}(1995)}]{nagaev1995influence}%
  \BibitemOpen
  \bibfield  {author} {\bibinfo {author} {\bibfnamefont {K.~E.}\ \bibnamefont
  {Nagaev}},\ }\bibfield  {title} {\bibinfo {title} {Influence of
  electron-electron scattering on shot noise in diffusive contacts},\ }\href
  {https://doi.org/10.1103/PhysRevB.52.4740} {\bibfield  {journal} {\bibinfo
  {journal} {Phys. Rev. B}\ }\textbf {\bibinfo {volume} {52}},\ \bibinfo
  {pages} {4740} (\bibinfo {year} {1995})}\BibitemShut {NoStop}%
\bibitem [{\citenamefont {Kozub}\ and\ \citenamefont
  {Rudin}(1995)}]{kozub1995shot}%
  \BibitemOpen
  \bibfield  {author} {\bibinfo {author} {\bibfnamefont {V.~I.}\ \bibnamefont
  {Kozub}}\ and\ \bibinfo {author} {\bibfnamefont {A.~M.}\ \bibnamefont
  {Rudin}},\ }\bibfield  {title} {\bibinfo {title} {Shot noise in mesoscopic
  diffusive conductors in the limit of strong electron-electron scattering},\
  }\href {https://doi.org/10.1103/PhysRevB.52.7853} {\bibfield  {journal}
  {\bibinfo  {journal} {Phys. Rev. B}\ }\textbf {\bibinfo {volume} {52}},\
  \bibinfo {pages} {7853} (\bibinfo {year} {1995})}\BibitemShut {NoStop}%
\bibitem [{\citenamefont {Hofmann}\ and\ \citenamefont
  {Gran}(2023)}]{PhysRevB.108.L121401}%
  \BibitemOpen
  \bibfield  {author} {\bibinfo {author} {\bibfnamefont {J.}~\bibnamefont
  {Hofmann}}\ and\ \bibinfo {author} {\bibfnamefont {U.}~\bibnamefont {Gran}},\
  }\bibfield  {title} {\bibinfo {title} {{Anomalously long lifetimes in
  two-dimensional Fermi liquids}},\ }\href
  {https://doi.org/10.1103/PhysRevB.108.L121401} {\bibfield  {journal}
  {\bibinfo  {journal} {Phys. Rev. B}\ }\textbf {\bibinfo {volume} {108}},\
  \bibinfo {pages} {L121401} (\bibinfo {year} {2023})}\BibitemShut {NoStop}%
\bibitem [{\citenamefont {Huang}\ \emph {et~al.}(2023)\citenamefont {Huang},
  \citenamefont {Farrell}, \citenamefont {Friedman}, \citenamefont {Zane},
  \citenamefont {Glorioso},\ and\ \citenamefont {Lucas}}]{lucas2023}%
  \BibitemOpen
  \bibfield  {author} {\bibinfo {author} {\bibfnamefont {X.}~\bibnamefont
  {Huang}}, \bibinfo {author} {\bibfnamefont {J.~H.}\ \bibnamefont {Farrell}},
  \bibinfo {author} {\bibfnamefont {A.~J.}\ \bibnamefont {Friedman}}, \bibinfo
  {author} {\bibfnamefont {I.}~\bibnamefont {Zane}}, \bibinfo {author}
  {\bibfnamefont {P.}~\bibnamefont {Glorioso}},\ and\ \bibinfo {author}
  {\bibfnamefont {A.}~\bibnamefont {Lucas}},\ }\bibfield  {title} {\bibinfo
  {title} {Generalized time-reversal symmetry and effective theories for
  nonequilibrium matter},\ }\href {https://arxiv.org/abs/2310.12233} {\
  (\bibinfo {year} {2023})},\ \Eprint {https://arxiv.org/abs/2310.12233}
  {arXiv:2310.12233 [cond-mat.stat-mech]} \BibitemShut {NoStop}%
\bibitem [{\citenamefont {Zhang}\ \emph {et~al.}(2026)\citenamefont {Zhang},
  \citenamefont {Samajdar},\ and\ \citenamefont
  {Gopalakrishnan}}]{zhang2026detecting}%
  \BibitemOpen
  \bibfield  {author} {\bibinfo {author} {\bibfnamefont {Y.~F.}\ \bibnamefont
  {Zhang}}, \bibinfo {author} {\bibfnamefont {R.}~\bibnamefont {Samajdar}},\
  and\ \bibinfo {author} {\bibfnamefont {S.}~\bibnamefont {Gopalakrishnan}},\
  }\bibfield  {title} {\bibinfo {title} {Detecting vortex motion through
  spatially correlated nonequilibrium noise},\ }\href
  {https://arxiv.org/abs/2605.18941} {\  (\bibinfo {year} {2026})},\ \Eprint
  {https://arxiv.org/abs/2605.18941} {arXiv:2605.18941 [cond-mat.supr-con]}
  \BibitemShut {NoStop}%
\bibitem [{\citenamefont {Kapitulnik}\ \emph {et~al.}(2019)\citenamefont
  {Kapitulnik}, \citenamefont {Kivelson},\ and\ \citenamefont
  {Spivak}}]{RevModPhys.91.011002}%
  \BibitemOpen
  \bibfield  {author} {\bibinfo {author} {\bibfnamefont {A.}~\bibnamefont
  {Kapitulnik}}, \bibinfo {author} {\bibfnamefont {S.~A.}\ \bibnamefont
  {Kivelson}},\ and\ \bibinfo {author} {\bibfnamefont {B.}~\bibnamefont
  {Spivak}},\ }\bibfield  {title} {\bibinfo {title} {Colloquium: Anomalous
  metals: Failed superconductors},\ }\href
  {https://doi.org/10.1103/RevModPhys.91.011002} {\bibfield  {journal}
  {\bibinfo  {journal} {Rev. Mod. Phys.}\ }\textbf {\bibinfo {volume} {91}},\
  \bibinfo {pages} {011002} (\bibinfo {year} {2019})}\BibitemShut {NoStop}%
\end{thebibliography}%

\end{document}